\documentclass[conference]{IEEEtran}
\IEEEoverridecommandlockouts

\usepackage[english]{babel}

\usepackage{cite}
\usepackage{hyperref}
\hypersetup{colorlinks=true, unicode=true, linkcolor=[rgb]{0.10,0.05,0.67}, citecolor=[rgb]{0.10,0.05,0.67}}
\usepackage{multirow}
\usepackage{booktabs}
\usepackage{makecell}
\usepackage{tikz}
\usetikzlibrary{positioning, shapes, arrows, arrows.meta}
\usepackage{tabularx}
\usepackage{enumitem}
\usepackage{ragged2e}
\usepackage{listings}

\lstdefinestyle{json}{
  basicstyle=\ttfamily\scriptsize,
  numbers=left,
  numberstyle=\tiny,
  stepnumber=1,
  numbersep=5pt,
  showstringspaces=false,
  breaklines=true,
  frame=single,
  rulecolor=\color{black},
  xleftmargin=10pt,
  moredelim=**[is][\color{blue}]{~}{~}
}
\newcolumntype{L}[1]{>{\RaggedRight\arraybackslash}p{#1}}

\begin{document}

\title{Trustworthiness in Digital Twin Systems: Systematic Review and Research Horizons}

\author{
\IEEEauthorblockN{Chi Fai David Lam, \ \ \ Aad van Moorsel, \ \ \ Zoya Pourmirza}
\IEEEauthorblockA{\textit{School of Computer Science} \\
\textit{University of Birmingham} \\
Birmingham, UK \\
\{c.f.d.lam,a.vanmoorsel,z.pourmirza\}@bham.ac.uk}
}

\maketitle

\begin{abstract}
  Digital Twins (DTs) are increasingly deployed across application domains, yet the treatment of trust-related issues remains unevenly addressed. To examine whether and how trust is discussed in the current landscape, we conducted a systematic review of existing DT review papers and a mapping of their abstracts. Seven trust-related challenges and seven trust-enhancing strategies were defined to guide the analysis, enabling the trust focus of each paper to be characterised.

  By aggregating the challenges and strategies referenced across domains, distinct patterns of emphasis were observed. With certain domains consistently sharing similar spectrum of trust concerns, four integration types, including human-centred, safety-critical, context-specific, and technologically-driven, were identified as emergent categories reflecting how trust is prioritised in different deployment contexts. Drawing on the characteristics of these types, several preliminary directions for future research were proposed. These include the development of trust-by-design principles to inform early-stage decision-making, the inclusion of trust metadata in platform schemas to prompt systematic developer consideration of trust factors, and the exploration of how architectural choices, such as federated DTs, influence user trust.
\end{abstract}

\section{Introduction}

Digital twins (DTs) have emerged as a key paradigm in modern digital systems \cite{Semeraro2021digital}, enabling continuous connectivity and data exchange between physical entities and their virtual representations \cite{Grieves2015digital}. DTs support real-time monitoring, analysis, and decision-making across a wide range of application domains \cite{Fuller2020digital}. However, the volume and velocity of data exchanged between physical and digital counterparts increases system vulnerability \cite{Jeremiah2024a}. A variety of issues, including cyberattacks, operational errors, and model or data inconsistencies, can cause anomalies that propagate to the physical system and affect users \cite{Suhail2023the}. As a result, concerns related to security, privacy, trust, governance, ethics, and resilience are receiving growing attention \cite{Kustelega2024privacy, vanDerAalst2021resilient}.

These concerns fall under the broader concept of trustworthiness. Philosophically, trustworthiness can be understood at two levels \cite{Kelp2023what}. Trustworthiness simpliciter captures the broad, general notion of meeting a baseline set of obligations expected of DTs across domains, including operating securely, reliably, transparently, ethically, and resiliently. Trustworthiness to $\phi$ refers to the ability to meet specific obligations that arise in particular contexts, such as strict safety requirements in healthcare or specialised resilience thresholds in critical national infrastructure. This distinction highlights that trustworthiness is not a binary property but depends on both general expectations for DT operation and the domain-specific tasks it is designed to support.

Trustworthiness is critical for the safe and sustained adoption of DTs, particularly in high-stakes or sensitive environments. Yet, these concerns are often addressed inconsistently across application domains, and solutions proposed in one area may not be readily transferable to others \cite{Jeremiah2024a}. This research aims to examine how such cross-cutting issues are discussed in existing DT review papers and to provide a structured perspective on emerging patterns of trust consideration, highlighting potential directions for future research.

The remainder of this paper is structured as follows. Section~\ref{sec:application_domains} introduces the concept of DTs and their application across different domains. Sections~\ref{sec:key-challenges} and \ref{sec:strategies} define the key trust-related challenges and potential strategies to address them, providing a conceptual framework for the review. Section~\ref{sec:methodology} outlines the systematic review and mapping approach, including data collection and analysis procedures. Section~\ref{sec:results} presents the findings, while Section~\ref{sec:discussions} discusses emerging patterns and categorises application domains based on trust considerations. Section~\ref{sec:future_work} highlights directions for future research, and Section~\ref{sec:conclusion} summarises the contributions of this study.

\begin{figure*}[htbp]
    \centering
    \includegraphics[width=\linewidth]{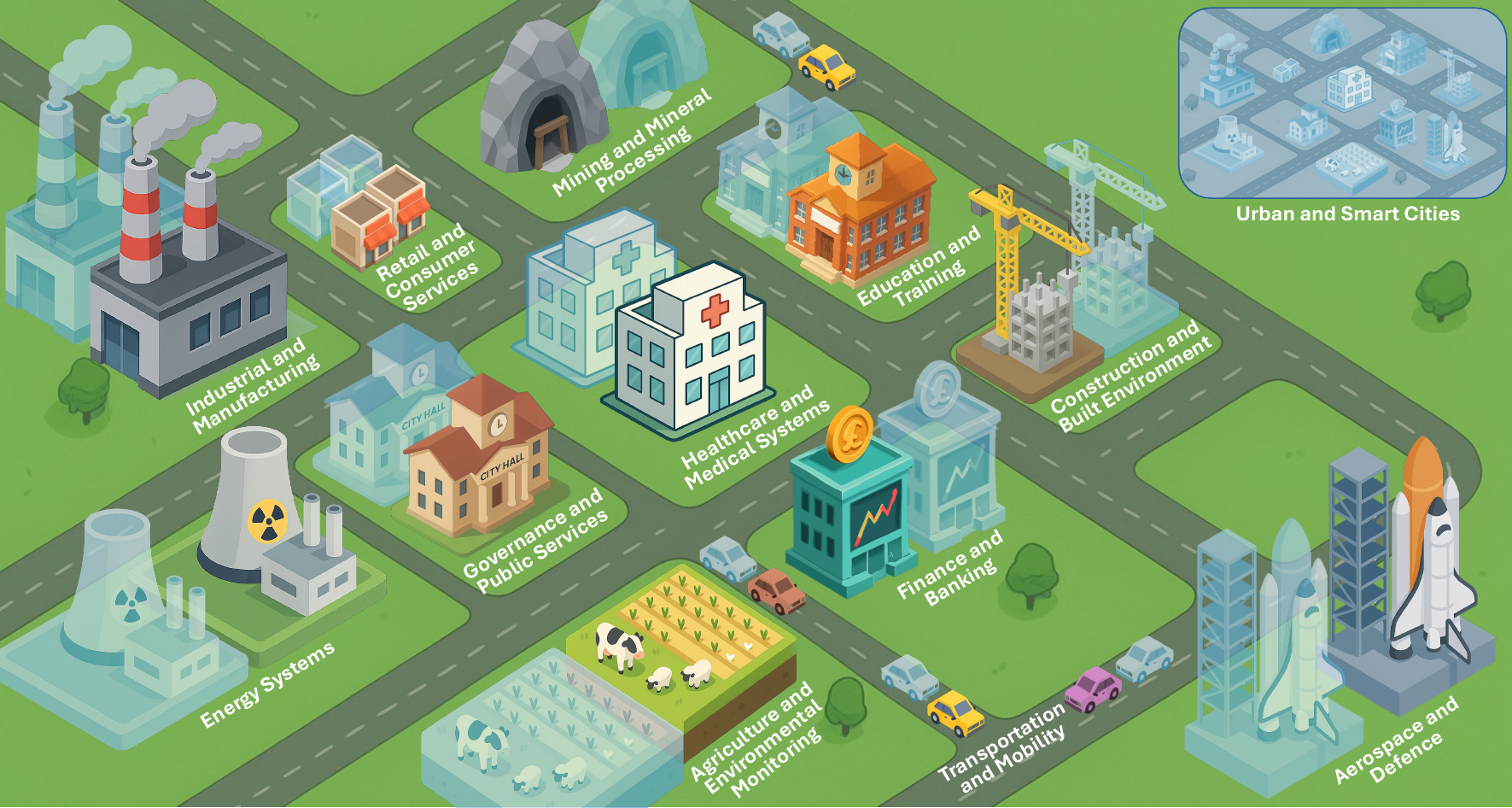}
    \caption{Thirteen major application domains, spanning physical, organisational, and social systems, including both micro- and macro-level social contexts.}
    \label{fig:application_areas}
\end{figure*}

\section{Concepts, Lifecycle, and Application Domains}
\label{sec:application_domains}

A DT builds on the idea of creating digital representations of physical systems. It evolved from Digital Models (DMs), which were static representations requiring manual updates, to Digital Shadows (DSs), where real-time sensor data enabled one-way updates from the physical system. A true Digital Twin goes further by enabling bidirectional, real-time data exchange, so changes in either the digital or physical domain are automatically mirrored, supporting monitoring, simulation, prediction, and even active control \cite{Kritzinger2018digital}. First introduced by NASA in 1970 for aerospace vehicle monitoring and mission planning \cite{Glaessgen2012digital}, DTs have since become a key enabler throughout the lifecycle of products and services, supporting activities from conceptualisation and virtual prototyping to testing, design optimisation, operational monitoring, and mission planning \cite{Rasheed2020digital}.

Across domains, DTs offer a common set of capabilities centred on real-time data integration, monitoring, simulation, and predictive analytics. By continuously synchronising physical systems with their digital counterparts, DTs support scenario exploration, fault detection, performance optimisation, and data-driven decision making. Although these foundational functions are broadly shared, each application domain uses DTs differently according to its operational requirements, constraints, and optimisation goals. Over time, DTs have also expanded beyond physical assets to represent intangible systems such as government processes, financial products, and other complex operations \cite{Duan2025the}. Their versatility and ongoing evolution continue to drive new applications. Figure~\ref{fig:application_areas} illustrates how major DT application domains span both micro-level systems and macro-level social contexts, covering physical, organisational, and societal domains. Table~\ref{tab:dt_domains} summarises these domains and highlights their core uses and domain-specific objectives.

\begin{table*}[htbp]
\centering
\scriptsize
\caption{Overview of thirteen major application domains, their modelling focus, data sources, objectives, and integration contexts.}
\label{tab:dt_domains}
\begin{tabular}{L{2.5cm} L{2.5cm} L{2.5cm} L{4cm} L{2.8cm}}
\toprule
\textbf{Domain} & \textbf{Model of} & \textbf{Data Source} & \textbf{Domain-Specific Objectives} & \textbf{Integration Context} \\
\midrule

Industrial and Manufacturing \cite{Bokhtiar2025digital, Guillaume2024digital} &
Machines, production lines, workflows &
Sensor data, production data, operational data, maintenance logs &
Design \& prototyping support, process optimisation, predictive maintenance, resource planning, safe human-robot interaction, sustainable lifecycle management &
Factory information systems, digital lifecycle platforms \\
\addlinespace

Mining and Mineral Processing \cite{Plavsic2023industrial, Bazi2023generic} &
Mining equipment, mine sites, material flows, extraction processes &
Equipment sensors, operational logs, environmental data &
Yield optimisation, safety assurance, predictive maintenance, asset condition monitoring &
Industrial monitoring platforms, cloud-based analytics \\
\addlinespace

Energy Systems \cite{Han2023cloud} &
Generation units, distributed resources, grid behaviour &
Real-time energy output, control data, historical performance &
Grid balancing, demand forecasting, energy optimisation, predictive maintenance &
Energy management platforms, edge/cloud systems \\
\addlinespace

Construction and Built Environment \cite{Mojtaba2025digital, Baghalzadeh2022internet} &
Buildings, construction sites, structural components &
Building models, environmental sensors, progress monitoring data &
Real-time coordination, design optimisation, safety management, cost control, construction monitoring &
Building modelling systems, sensor networks, planning tools \\
\addlinespace

Urban and Smart Cities \cite{Deng2021systematic} &
City infrastructure, mobility networks, utilities, public services &
City-scale sensor networks, infrastructure monitoring data, mobility data &
Urban planning, service optimisation, environmental monitoring, disaster preparedness, climate resilience &
Smart-city platforms, IoT sensor ecosystems \\
\addlinespace

Transportation and Mobility \cite{Irfan2024toward} &
Vehicles, transport networks, traffic systems &
Vehicle telemetry, roadside sensors, mobility data &
Traffic optimisation, scenario testing, predictive maintenance, safety enhancement, congestion reduction &
Transport management platforms, connected infrastructure \\
\addlinespace

Aerospace and Defence \cite{Yin2020application, Mendi2022digital} &
Aircraft, spacecraft, defence systems &
Structural sensors, flight data, mission telemetry, operational logs &
Fault diagnosis, predictive maintenance, mission planning, operational optimisation, enhance reliability &
High-fidelity simulation systems, secure communication networks \\
\addlinespace

Healthcare and Medical Systems \cite{Sadee2025medical} &
Patients, clinical environments, physiological systems, clinical workflows &
Health records, medical imaging, wearable device data &
Personalised medicine, treatment planning, early diagnosis, reduction of unnecessary interventions, resource planning &
Clinical decision-support systems, integrated health records \\
\addlinespace

Agriculture and Environmental Monitoring \cite{Stefano2023smart} &
Crops, livestock, soil systems, ecosystems &
Soil \& crop sensors, weather data, environmental monitoring &
Yield forecasting, resource optimisation, pest management, sustainability tracking &
Agricultural management platforms, environmental monitoring tools \\
\addlinespace

Education and Training \cite{Dimitris2023hybrid} &
Learning environments, training processes &
Learner activity data, simulation interactions, performance data &
Personalised learning, adaptive instruction, skill assessment, immersive training &
Digital learning platforms, simulation systems \\
\addlinespace

Retail and Consumer Services \cite{Kumpel2021semantic} &
Stores, supply chains, customer environments &
Product tracking data, customer behaviour data, inventory data &
Inventory optimisation, customer engagement, logistics efficiency, strategic decision-making &
Retail management systems, analytics platforms \\
\addlinespace

Governance and Public Services \cite{Eom2022the, Anshari2022enhancing} &
Administrative processes, public infrastructure, community systems &
Government records, service usage patterns, infrastructure data &
Policy planning, decision-support, service optimisation, crisis response, value co-creation &
Public sector digital platforms, citizen service portals \\
\addlinespace

Finance and Banking \cite{Anshari2022digital}  &
Financial products, user behaviour, market systems &
Transaction histories, behavioural analytics, market data &
Risk assessment, personalised planning, fraud detection, portfolio optimisation, compliance monitoring &
Financial analytics systems, digital banking services \\
\bottomrule
\end{tabular}
\end{table*}

\section{Key Challenges and Concerns}
\label{sec:key-challenges}

As DTs are increasingly deployed in real-world contexts, ensuring the trustworthiness of the system so that users and stakeholders can have confidence in it has become a critical concern. One established way to understand trustworthiness is through the concept of dependability, which describes a system's ability to deliver correct and predictable service under both normal and exceptional conditions. Dependability includes several key attributes: availability refers to the system being ready to provide service; reliability concerns the continuous delivery of correct service; safety involves avoiding harmful or catastrophic outcomes; integrity is the protection against unauthorised or improper modifications; maintainability relates to the system's capacity to be updated and repaired; and confidentiality involves the protection of sensitive information from unauthorised disclosure \cite{Avizienis2004basic}.

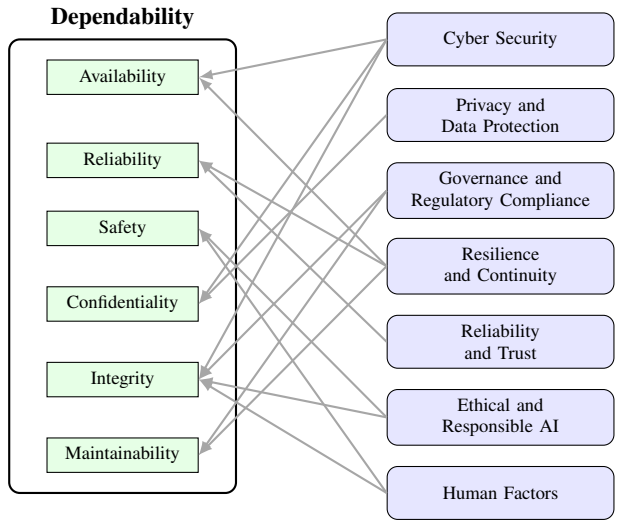
\begin{figure}[!ht]
\centering
\begin{tikzpicture}[
    challenge/.style={rectangle, draw, rounded corners, align=center, fill=blue!10, font=\scriptsize, minimum width=3cm, minimum height=0.7cm},
    dependability/.style={rectangle, draw, align=center, fill=green!10, font=\scriptsize, minimum width=2cm},
    edge/.style={-latex, thick, draw=gray!70}
]

\node[dependability] (availability) at (0,5.5) {Availability};
\node[dependability] (reliability) at (0,4.4) {Reliability};
\node[dependability] (safety) at (0,3.5) {Safety};
\node[dependability] (confidentiality) at (0,2.5) {Confidentiality};
\node[dependability] (integrity) at (0,1.5) {Integrity};
\node[dependability] (maintainability) at (0,0.5) {Maintainability};

\node[draw, rounded corners, thick, 
      label=above:{\small \bfseries Dependability}, 
      minimum width=3cm, minimum height=6cm] 
      (dependabilityBox) at (0, 3) {};

\node[challenge] (cyber) at (5,6) {Cyber Security};
\node[challenge] (privacy) at (5,5) {Privacy and\\Data Protection};
\node[challenge] (governance) at (5,4) {Governance and\\Regulatory Compliance};
\node[challenge] (resilience) at (5,3) {Resilience\\and Continuity};
\node[challenge] (trust) at (5,2) {Reliability\\and Trust};
\node[challenge] (ethical) at (5,1) {Ethical and\\Responsible AI};
\node[challenge] (human) at (5,0) {Human Factors};

\draw[edge] (cyber.west) -- (availability.east);
\draw[edge] (cyber.west) -- (integrity.east);
\draw[edge] (cyber.west) -- (confidentiality.east);
\draw[edge] (privacy.west) -- (confidentiality.east);
\draw[edge] (governance.west) -- (maintainability.east);
\draw[edge] (governance.west) -- (integrity.east);
\draw[edge] (resilience.west) -- (availability.east);
\draw[edge] (resilience.west) -- (reliability.east);
\draw[edge] (resilience.west) -- (maintainability.east);
\draw[edge] (trust.west) -- (reliability.east);
\draw[edge] (ethical.west) -- (safety.east);
\draw[edge] (ethical.west) -- (integrity.east);
\draw[edge] (human.west) -- (safety.east);
\draw[edge] (human.west) -- (integrity.east);

\end{tikzpicture}
\caption{Bipartite mapping of challenges to dependability attributes. Each edge indicates which dependability attribute(s) a given challenge impacts.}
\label{fig:dependability_bipartite}
\end{figure}

These dependability attributes align closely with the key challenges facing DT deployment (see Figure~\ref{fig:dependability_bipartite}). Cyber security addresses threats to integrity, availability, and confidentiality by protecting DTs against malicious attacks and unauthorised access. Privacy and data protection ensure that personal and sensitive information is handled appropriately, reinforcing confidentiality and fostering user trust. Governance and regulatory compliance support maintainability by establishing clear structures for oversight, adaptation, and accountability, while also promoting integrity through adherence to standards. Resilience enhances availability, reliability, and maintainability by enabling DTs to continue operating under stress or recover from failures. Reliability and trust are strengthened by consistent system behaviour and credible outputs over time. Ethical and responsible AI practices advance safety by reducing the risks of bias, harm, or unintended consequences, while also supporting integrity through fairness and transparency. Finally, human factors impact safety and integrity, as errors, misinterpretation, or poor interaction design can compromise correct operation and introduce risks.

This framing highlights how the challenges of DT implementation can be understood in terms of the core attributes of dependability, linking practical concerns in real-world deployment to foundational system qualities. The following sections provide an overview of each challenge, its relevance to DTs, and key research themes addressing them.

It is important to note that this discussion focuses on challenges that affect trust and confidence. Fundamental technical and operational difficulties, such as high development and operational costs, computational and energy requirements, heterogeneous data environments, real-time data processing, model accuracy or complexity, and network constraints (latency and bandwidth), are not the primary focus here, even though they remain critical for the practical implementation of DT systems.

\subsection{Cyber Security}

The tight coupling of DTs with physical assets, cloud platforms, and heterogeneous communication networks significantly enlarges their attack surface, creating conditions in which conventional cyber-threats have heightened consequences. Although risks such as  distributed denial-of-service (DDoS) attacks, malware injection, privilege escalation, and man-in-the-middle exploits are not unique to DTs, their impact is amplified because DT operations rely on continuous, high-integrity data flows. Disruptions or manipulations at any point, whether through vulnerable communication channels, insecure data storage, or poorly protected IoT and edge devices, can compromise the fidelity of the digital representation, enabling data tampering, intellectual property theft, and unauthorised control \cite{Frustaci2018evaluating}. Compromised endpoints often function as pivot points for wider system breaches, exposing systemic rather than isolated weaknesses. Additionally, software-lifecycle risks, including malicious updates, version mismatches, and unpatched components, further illustrate the range of technical vulnerabilities that threaten DT integrity \cite{Behrang2019an}. Collectively, these interdependent risks challenge the assumption that DTs can operate as secure and trustworthy mirrors of physical systems, thereby undermining user confidence.

\subsection{Privacy and Data Protection}

As DTs increasingly process continuous streams of data not only between devices but also from users and environments, responsible data handling becomes essential. Personal or sensitive information such as personally identifiable information, behavioural patterns, or biometric data pose significant privacy risks if inadequately protected. Failures in anonymisation, consent management, or inconsistent privacy safeguards during data exchange and aggregation across multiple systems can lead to data exposure or enable cross-system linkage attacks that reidentify individuals even in anonymised datasets. These risks undermine user trust and raise serious ethical and legal concerns. Privacy challenges are especially acute in domains such as healthcare, smart cities, and personalised services, where sensitive and context-rich data are prevalent \cite{Mostert2016big, Kustelega2024privacy}. Such concerns underscore the growing importance of organisational accountability and adherence to evolving regulatory frameworks, which are addressed in the following section.

\subsection{Governance and Regulatory Compliance}

As DT ecosystems grow more complex, integrating heterogeneous components across organisational and sectoral boundaries introduces significant governance risks. A major challenge is the lack of widely accepted interoperability standards and unified governance models, resulting in unclear responsibilities, conflicting policies, and gaps in accountability as heterogeneous DTs become interconnected. Data governance issues, including inconsistent access control, insufficient audit mechanisms that hinder traceability, and ambiguous data ownership, especially in cross-organisational and public-private contexts, which further heighten the risk of misuse and disputes. These weaknesses also undermine the reliability of decision-making processes that depend on DT outputs. Regulatory compliance adds another layer of complexity. Frameworks such as GDPR, ENISA guidelines, ISO IEC 27001, and sector-specific rules like HIPAA are fragmented globally and subject to varying interpretations, leaving organisations uncertain about how to operate safely across jurisdictions \cite{Kumas2024international, Jorgensen2025digital}. The ongoing evolution and expansion of DT systems exacerbate the challenge of maintaining compliance over time. Collectively, these governance complexities affect scalability, legal standing, operational reliability, and stakeholder trust.

\subsection{Resilience and Continuity}

DTs depend on complex interactions among physical assets, digital components, and communication networks, making resilience and continuity essential yet difficult to maintain. Intrusion detection and prevention systems (IDPS) often fail to promptly identify or mitigate cyber threats, leaving DTs vulnerable to disruption \cite{Sayghe2025digital}. Many systems lack redundancy and failover mechanisms, causing downtime during failures. Resource constrained IoT sensors and edge devices with limited computational capacity weaken overall detection and recovery capabilities. Network outages or sensor malfunctions can disrupt synchronisation between physical assets and their digital counterparts, reducing accuracy and reliability. Delays in incident response can worsen the effects of security breaches or system failures, while limited system flexibility hampers adaptation to evolving threats. The interconnected nature of DT ecosystems increases the risk of cascading failures, complicating detection and containment of errors, particularly in large-scale environments like smart cities or industrial settings \cite{Berger2021survey}. Beyond operational consequences, these resilience challenges have direct implications for trust. If DTs cannot guarantee continuous, reliable, and accurate operation, users may lose confidence in the insights they provide. Ensuring resilience is therefore not only a technical requirement but also demands robust strategies to maintain reliable operations under dynamic conditions.

\subsection{Reliability and Trust}

Reliable, accurate, and timely data synchronisation between physical and digital components is fundamental to DT effectiveness. Data loss, degradation, or desynchronisation can lead to incorrect outputs, jeopardising safety and performance, especially in dynamic or time-sensitive environments where errors can cascade through interconnected systems \cite{Wael2024synchronization}. The integration of AI and machine learning adds complexity, as the opacity and evolving behaviour of these models make verification and interpretation difficult, particularly in continuously learning systems where unexpected changes may go unnoticed \cite{Kazuma2024explainable}. Trust also depends on verifying the authenticity and integrity of DT instances. Identity-related threats such as spoofing, impersonation, or rogue twin deployment risk injecting false data or unauthorised actions. AI-specific vulnerabilities, including adversarial manipulation and insider tampering targeting model components, further threaten predictability and stability, with potentially serious real-world consequences in critical domains \cite{Wang2023survey}. Maintaining consistency and transparency over time is essential to foster stakeholder confidence in DT systems.

\subsection{Ethical and Responsible AI}

As DTs increasingly leverage AI and machine learning for automated decision making, ensuring ethical and responsible behaviour is paramount. Algorithmic fairness remains a major challenge, particularly when models are trained on biased or unrepresentative data, risking skewed outcomes or unequal treatment across demographic groups \cite{Giovanola2023beyond}. The opacity of many AI models restricts transparency and accountability, especially where decisions impact individuals or communities \cite{Kazuma2024explainable}. Responsible data use is equally critical, as operational data collection may inadvertently capture sensitive behavioural or environmental information, leading to unauthorised profiling beyond users' original consent \cite{Lee2022toward}. Without clear ethical norms and guidelines, the interaction between biased models, opaque decision processes, and sensitive data can systematically undermine stakeholder trust and provoke societal resistance. Ensuring ethical and responsible AI therefore requires ongoing critical reflection, governance, and engagement with broader social implications, rather than reliance on purely technical fixes.

\subsection{Human Factors}

Human involvement is integral to every stage of the DT lifecycle, making human factors crucial for the success and reliability of the system. The "human in the loop" approach facilitates expert oversight and intervention, thereby enhancing adaptability and contextual understanding \cite{Ashwin2023digital}. However, it also introduces risks such as misinterpreting DT outputs, responding too slowly, and taking the wrong action. These risks can be exacerbated by cognitive overload, insufficient training and unclear interfaces, potentially leading to operational failures or safety incidents \cite{Horvath2023investigating}. Humans also pose a cyber security vulnerability through actions such as phishing or introducing malware \cite{Pollini2021leveraging}. Balancing automation with meaningful human control is challenging, as overreliance on automation can erode situational awareness, while excessive manual intervention may slow response times and reduce reliability. Effective DT design must therefore prioritise human-machine collaboration, interface clarity, and error resilience, recognising that human factors are not peripheral but central to operational integrity and stakeholder trust.

\section{Mitigation Strategies for Trustworthiness}
\label{sec:strategies}

Enhancing the trustworthiness of DTs requires a comprehensive approach that tackles technical, organisational and human-centred issues in a coordinated and integrated way. Due to the complexity and interdependence of these issues, effective solutions rarely target a single aspect, instead spanning multiple overlapping domains of research and practice. This section outlines the key dimensions of solutions that collectively support the creation of secure, reliable and ethically responsible DT systems.

\begin{figure}[ht]
    \centering
    \includegraphics[width=9cm]{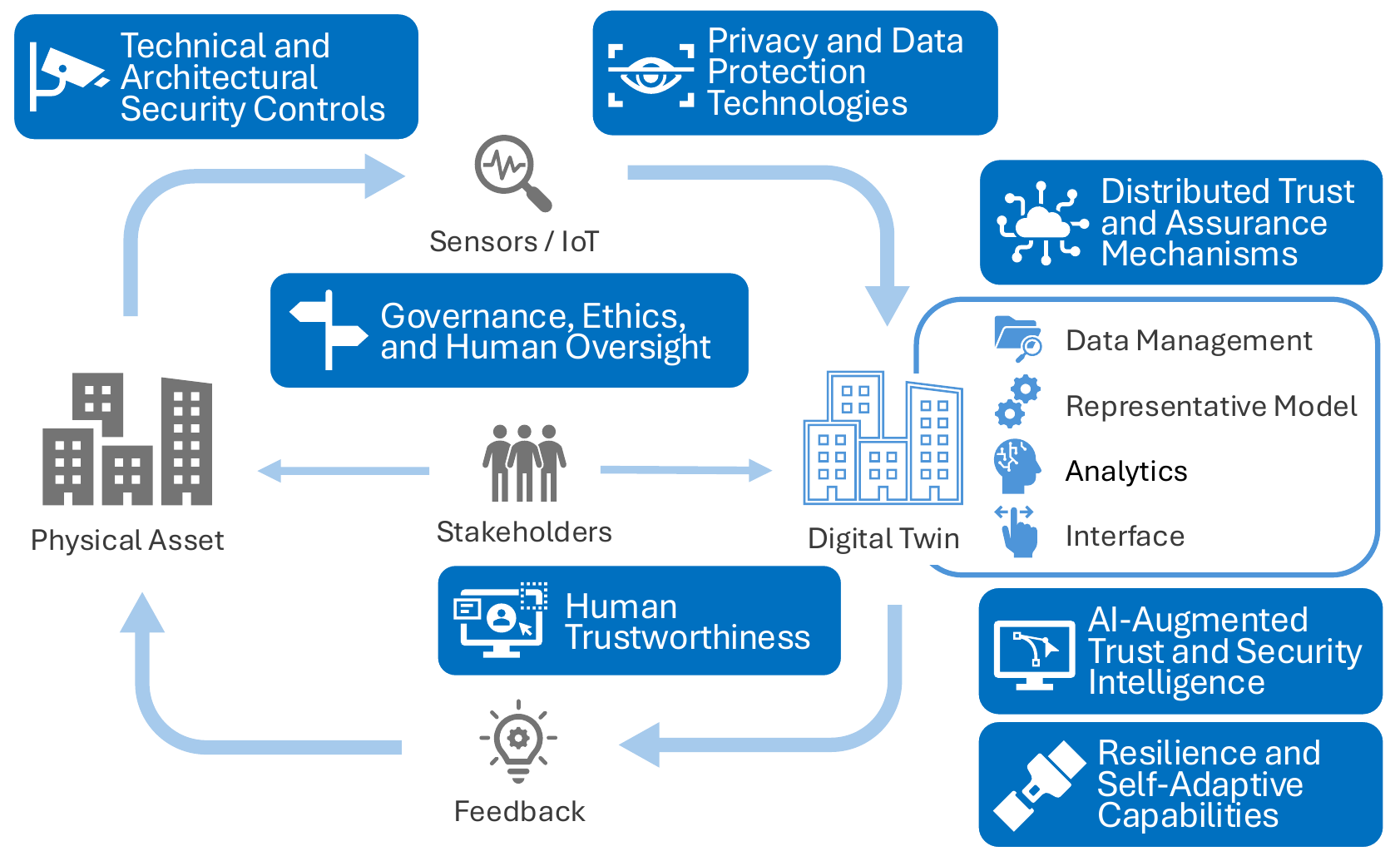}
    \caption{Seven key mitigation strategies to enhance DT trustworthiness.}
    \label{fig:strategies_span}
\end{figure}

These dimensions (see Figure~\ref{fig:strategies_span}) span the full lifecycle of a DT. They include technical and architectural security controls that provide a foundation for protection against cyber threats, as well as privacy-enhancing technologies that are designed to safeguard sensitive data. AI-driven methods facilitate intelligent threat detection, predictive maintenance and transparent decision-making processes. At the organisational level, governance frameworks and regulatory compliance ensure accountability, and human-centred design principles promote usability, interpretability and trust. Furthermore, resilience strategies enable systems to withstand and recover from disruptions, while distributed trust mechanisms, such as auditability and decentralised verification, provide transparency and assurance across DT ecosystems.

Together, these dimensions reflect the multidisciplinary efforts that are essential for the safe, effective and responsible deployment of DTs in real-world settings.

\subsection{Technical and Architectural Security Controls}

Robust technical and architectural controls are essential for securing DT systems and maintaining their integrity throughout the entire digital and physical component stack. As DTs operate in cloud, edge and on-device environments, it is paramount that data transmission and system interactions are secured. Secure communication protocols such as TLS and MQTT-S protect data flows between sensors, edge devices and cloud platforms. End-to-end encryption preserves the confidentiality and integrity of data in transit and at rest \cite{Frustaci2018evaluating}. Implementing a zero trust architecture further strengthens DT ecosystems by enforcing continuous authentication, minimising implicit trust and reducing the risk of adversaries moving laterally across the system \cite{Samaniego2018zero}.

At the hardware level, secure boot mechanisms and trust anchors, including Trusted Platform Modules (TPMs) and Hardware Security Modules (HSMs), are used to establish device integrity, which is particularly crucial in IoT-intensive deployments \cite{Frustaci2018evaluating}. To counter evolving threats, DT networks must integrate intrusion detection and prevention systems (IDPS) tailored to cyber-physical environments, as well as runtime monitoring capable of identifying anomalies and unexpected behaviours \cite{Mitchell2014survey}. System reliability also hinges on the prompt delivery of secure software updates to address vulnerabilities and prevent the exploitation of outdated components.

Identity and access management (IAM) frameworks are essential for controlling access across distributed DT environments. These frameworks are bolstered by secure APIs and interface protections that mitigate the risks associated with interoperability and data exchange. Due to their limited processing capabilities and security features, edge and IoT devices require dedicated hardening strategies to minimise their vulnerability as potential attack vectors \cite{Talluri2022identity}. Together, these measures create a robust DT architecture that can withstand various threats while ensuring secure, uninterrupted operation within complex, interconnected systems.

\subsection{Privacy and Data Protection Technologies}

As DT systems increasingly rely on the continuous collection of data from physical environments, devices and users, the protection of personal and sensitive information is becoming a key aspect of trustworthiness. Privacy-enhancing technologies provide essential mechanisms for managing risk and ensuring the responsible use of data throughout the DT lifecycle \cite{Rolf2010internet}. Techniques such as differential privacy limit the reidentification of individuals when sensor and operational data are shared, while anonymisation and pseudonymisation reduce identifiability during storage and processing \cite{Husnoo2021differential}.

In distributed DT deployments, federated learning enables collaborative model training without centralising raw data, thus preserving local privacy while supporting system-wide intelligence \cite{Viraaji2021Survey}. For high-stakes or cross-organisational collaborations, secure multi-party computation (SMPC) and homomorphic encryption provide cryptographic assurances, enabling analytics to be performed without exposing the underlying data \cite{Li2025secure}. Context-aware consent management systems bolster individual autonomy by aligning data use with situational and regulatory expectations \cite{Lee2022toward}.

The privacy-by-design principles of data minimisation and purpose limitation are crucial for limiting unnecessary data exposure and reducing long-term risk. Privacy threat modelling identifies and addresses vulnerabilities specific to data processing architectures by accounting for multi-source data flows and complex system interactions \cite{Azam2023data}. To promote accountability, privacy risk assessments and compliance audits evaluate adherence to legal frameworks and internal policies \cite{Damjanovic2018digital}. Finally, a secure data lifecycle management approach encompasses responsible collection, secure processing and timely deletion to maintain privacy protections throughout \cite{Kumar2020data}. Together, these technologies establish a proactive, layered privacy strategy that enables DT systems to function effectively while respecting data protection obligations and societal expectations.

\subsection{AI-Augmented Trust and Security Intelligence}

As DT systems grow in scale and complexity, artificial intelligence (AI) is playing an increasingly pivotal role in enhancing their trustworthiness, security and resilience. Machine learning-based anomaly detection enables the real-time monitoring of DT behaviours and the identification of deviations that may indicate failures or cyber threats. AI-driven threat prediction and risk scoring models allow for the proactive assessment of vulnerabilities and pre-emptive mitigation \cite{Xu2023digital, Samuel2025artificial}. Reinforcement learning shows promise in dynamically adjusting access control policies in response to contextual changes, thereby improving responsiveness within fluid DT environments \cite{Jiang2022beyond}.

AI can also enhance system visibility by supporting asset classification and vulnerability prioritisation, helping organisations to focus their security efforts effectively. AI-driven predictive maintenance correlates operational data with security signals to reduce downtime and exposure to risk \cite{Sohel2025aidriven}. In high-stakes applications involving automated decision-making, it is essential to safeguard AI models against adversarial manipulation \cite{Lu2023strategic}. Explainable AI (XAI) methods address the opacity of complex models by providing human-understandable justifications for system actions, thereby improving transparency \cite{Kazuma2024explainable}.

Behavioural analytics, powered by AI, detect insider threats or misuse by identifying abnormal usage patterns \cite{Dommari2021ai}. AI-enhanced situational awareness tools synthesise vast amounts of telemetry, event and contextual data to provide human operators with actionable insights \cite{Wu2024security}. Furthermore, adaptive security policy generation allows DT systems to evolve their defences in response to changing threat landscapes \cite{Okegbile2025artificial}. When these AI-driven techniques are effectively integrated, they foster a more intelligent, responsive and transparent DT ecosystem, thereby bolstering user trust and operational integrity.

\subsection{Governance, Ethics, and Human Oversight}

Trust in DT systems encompasses more than just technical measures; it also includes governance models, ethical safeguards and user-centred design principles that guide their deployment and operation. Policy-aware architectures ensure that DT functionality aligns with evolving regulatory requirements, thereby embedding compliance within the technical design \cite{Kassem2023adaptive}. Cross-sector governance frameworks, which are often established through public-private partnerships, facilitate coordinated oversight in large-scale or multi-stakeholder contexts \cite{Sawsan2024collaboration}.

In areas affecting human well-being, such as healthcare, mobility or urban management, ethical guidelines for AI-enabled decision-making help to prevent unintended harm, bias or discrimination \cite{Ferrell2024theoretical}. Transparency and auditability mechanisms are essential for accountability, especially when opaque or automated systems affect rights or safety. This involves maintaining clear records of decision-making processes and ensuring the traceability of data sources and processing logic \cite{Kroll2021outlining}.

Involving stakeholders meaningfully in DT design and validation enhances legitimacy, ensures diverse perspectives are considered and aligns systems with user values \cite{Nirnaya2024stakeholders}. Human-in-the-loop control models are essential in safety-critical situations, enabling human oversight to mitigate the risks of autonomous failure or inappropriate system behaviour \cite{Ashwin2023digital}.

In public or shared environments such as smart cities, digital ethics play a key role in data governance, particularly with regard to issues such as fairness, surveillance and civic accountability \cite{Kitchin2016the}. Mechanisms such as context-aware consent management, data minimisation and data trusts promote equitable and transparent control over personal and community information \cite{Lee2022toward, Rahma2022survey}. Managing user trust also involves setting clear expectations about the capabilities and limitations of DT. Finally, clearly defined organisational roles and responsibilities ensure sustainable governance by delineating accountability for system operation, oversight and incident response \cite{Nirnaya2024stakeholders}. Together, these practices establish a governance and ethical framework that enables DTs to operate in a responsible and inclusive manner that aligns with societal values.

\subsection{Resilience and Self-Adaptive Capabilities}

Resilient distributed systems require architectures that can maintain continuous operation in the face of faults, disruptions or attacks. Self-healing designs enable DT infrastructures to detect and recover from failures autonomously, thereby minimising downtime and preserving integrity \cite{Kambala2024intelligent}. Redundancy and failover mechanisms across network, hardware, and software layers provide backup resources to ensure that critical functions persist during outages \cite{Rullo2019redundancy}.

Synchronisation and recovery mechanisms maintain alignment between physical assets and their digital counterparts, which is crucial for ensuring the accuracy of data and the reliability of operations \cite{Wang2021synchronization}. Real-time fault detection and mitigation tools swiftly identify anomalies and initiate corrective action before errors escalate \cite{Stoumpos2021novel}. Dynamic reconfiguration enables DT systems to adapt their operational modes or resource allocation in response to evolving threats or faults, thereby enhancing their flexibility and survivability \cite{Leng2020digital}.

Cyber-resilience assessment and benchmarking frameworks offer measurable metrics to systematically evaluate these capabilities, supporting continuous improvement \cite{Haque2018cyber}. Chaos engineering, adapted for DT contexts, deliberately introduces faults to test robustness and expose hidden vulnerabilities. Understanding the complex interdependencies within DT ecosystems is essential for anticipating and preventing cascading failures that could amplify the impact of an incident \cite{Fogli2024chaos}.

Comprehensive incident response and disaster recovery planning prepares organisations to contain and remediate disruptions efficiently, thereby preserving stakeholder confidence and ensuring operational continuity \cite{Farok2024incident}. Together, these resilience and self-adaptive strategies ensure that DT systems can operate reliably in dynamic and unpredictable environments, thereby reinforcing their trustworthiness and long-term viability.

\subsection{Distributed Trust and Assurance Mechanisms}

Trust and integrity within distributed DT ecosystems depend on robust mechanisms that provide transparency, verifiability and secure coordination among diverse participants. Blockchain technology provides a fundamental approach to this, offering secure data provenance and immutable audit trails that are essential for verifying the authenticity and history of distributed ledger (DL) data exchanges \cite{Dietz2019distributed}. Smart contracts automate access control and compliance enforcement, reducing the need for manual oversight and enhancing operational efficiency \cite{Iakovos2024secure}.

Verifiable computing techniques ensure that the computations and analytics of complex DT models produce correct, tamper-proof results, thereby bolstering confidence in automated decision-making \cite{Yu2017survey}. Decentralised identity and credential management systems establish trustworthy identities for DT entities, enabling secure interactions within federated or collaborative environments \cite{Mazzocca2025survey}. Trusted execution environments (TEEs) provide hardware-enforced secure computation zones that protect sensitive data and operations from insider threats and external attacks \cite{Xu2024survey}.

Consensus mechanisms facilitate coordinated agreement on data states and updates across distributed DT networks, supporting consistency and resilience without centralised control. Distributed logging and tamper-evident recordkeeping complement these methods by preserving detailed, verifiable histories of system events and transactions. Proof-of-integrity schemes validate the accuracy and completeness of DT models, addressing concerns about model drift or manipulation \cite{Sasikumar2023blockchain}.

Trust scoring and reputation systems assess the reliability of DT components and participants dynamically, guiding access decisions and fostering accountability \cite{Flavia2023reinforcement}. Token-based access control models introduce flexible, cryptographically secured permissioning that is tailored to decentralised platforms \cite{Wang2023smart}. Together, these distributed trust and assurance mechanisms form a comprehensive framework that underpins the secure, transparent and reliable operation of interconnected distributed ledger ecosystems.

\subsection{Human Trustworthiness}

Human actors play an equally important role in the operation and governance of DT systems, and their actions have a significant influence on the trustworthiness of these systems. Fostering awareness, accountability, and competence through structured education, organisational culture, and user-centred design enhances human trustworthiness.

Security awareness training and ethical training equip individuals to identify risks and behave responsibly in complex DT environments. Continuous capacity building ensures that stakeholders remain informed about emerging threats, technologies and best practices relevant to their roles \cite{Horvath2023investigating, Pollini2021leveraging}.

Human-centric design principles minimise the risk of user error by aligning interfaces and workflows with cognitive capabilities and decision-making processes \cite{Asad2023human}. Effective usability and interface design support situational awareness, reduce cognitive overload and facilitate intuitive interaction with complex DT systems \cite{Makarov2019the}. At the same time, a culture of shared responsibility is fostered through clear policies, strong leadership commitment and effective feedback mechanisms, which reinforce secure and ethical conduct \cite{Crowther2024blending}.

By systematically addressing human factors, organisations can strengthen the integrity, reliability and ethical operation of DT systems, ensuring that human involvement supports rather than undermines digital trust.

\bigskip
To provide a comprehensive overview of how the identified strategies contribute to enhancing DT trustworthiness, Table~\ref{tab:strategiesMapping} maps each strategy to the corresponding challenges it addresses. This mapping enables a systematic understanding of how the approaches collectively strengthen trustworthiness across multiple dimensions.

\setcellgapes{1.8pt}
\makegapedcells

\begin{table*}[htbp]
\centering
\caption{Mitigation strategies for enhancing DT trustworthiness mapped to the challenges they address. Symbols denote the extent of applicability: $\bullet$ for directly targeted challenges, and $\circ$ for strategies that help reduce the likelihood of the challenge}
\label{tab:strategiesMapping}
\scriptsize
\begin{tabular}{|l|l|*{30}{c|}}
\hline
\multicolumn{2}{|c|}{\textbf{Mitigation Strategy / Challenge}} &
\multicolumn{4}{c|}{\makecell[c]{\textbf{Cyber}\\\textbf{Security}}} &
\multicolumn{4}{c|}{\makecell[c]{\textbf{Privacy \&}\\\textbf{Data}\\\textbf{Protection}}} &
\multicolumn{5}{c|}{\makecell[c]{\textbf{Governance \&}\\\textbf{Regulatory}\\\textbf{Compliance}}} &
\multicolumn{5}{c|}{\makecell[c]{\textbf{Resilience \&}\\\textbf{Continuity}}} &
\multicolumn{4}{c|}{\makecell[c]{\textbf{Reliability}\\\textbf{and Trust}}} &
\multicolumn{4}{c|}{\makecell[c]{\textbf{Ethical \&}\\\textbf{Responsible}\\\textbf{AI}}} &
\multicolumn{4}{c|}{\makecell[c]{\textbf{Human}\\\textbf{Factors}}} \\
\cline{3-32}
\multicolumn{2}{|c|}{} &
\makebox[0pt][c]{\rotatebox{90}{Common attack vectors}} &
\makebox[0pt][c]{\rotatebox{90}{Vulnerable devices \& channels}} &
\makebox[0pt][c]{\rotatebox{90}{Software lifecycle risks}} &
\makebox[0pt][c]{\rotatebox{90}{Data \& asset compromise}} &
\makebox[0pt][c]{\rotatebox{90}{Sensitive data exposure}} &
\makebox[0pt][c]{\rotatebox{90}{Weak anonymisation methods}} &
\makebox[0pt][c]{\rotatebox{90}{Inadequate consent management}} &
\makebox[0pt][c]{\rotatebox{90}{Cross-system linkage risk}} &
\makebox[0pt][c]{\rotatebox{90}{Lack of interoperability standards}} &
\makebox[0pt][c]{\rotatebox{90}{Ambiguous roles \& accountability}} &
\makebox[0pt][c]{\rotatebox{90}{Inconsistent data governance}} &
\makebox[0pt][c]{\rotatebox{90}{Fragmented global regulations}} &
\makebox[0pt][c]{\rotatebox{90}{Evolving compliance complexity}} &
\makebox[0pt][c]{\rotatebox{90}{Ineffective threat detection}} &
\makebox[0pt][c]{\rotatebox{90}{Lack of redundancy \& failover}} &
\makebox[0pt][c]{\rotatebox{90}{Resource-constrained edge devices}} &
\makebox[0pt][c]{\rotatebox{90}{Disrupted asset synchronisation}} &
\makebox[0pt][c]{\rotatebox{90}{Risk of cascading failures}} &
\makebox[0pt][c]{\rotatebox{90}{Data loss \& desynchronisation}} &
\makebox[0pt][c]{\rotatebox{90}{AI model opacity \& complexity}} &
\makebox[0pt][c]{\rotatebox{90}{Identity spoofing \& impersonation}} &
\makebox[0pt][c]{\rotatebox{90}{AI adversarial \& insider risks}} &
\makebox[0pt][c]{\rotatebox{90}{Algorithmic bias \& unfairness}} &
\makebox[0pt][c]{\rotatebox{90}{Lack of model transparency}} &
\makebox[0pt][c]{\rotatebox{90}{Unauthorised data profiling}} &
\makebox[0pt][c]{\rotatebox{90}{Undefined ethical guidelines}} &
\makebox[0pt][c]{\rotatebox{90}{Misinterpretation of outputs}} &
\makebox[0pt][c]{\rotatebox{90}{Cognitive overload \& errors}} &
\makebox[0pt][c]{\rotatebox{90}{Insufficient training \& support}} &
\makebox[0pt][c]{\rotatebox{90}{Human-induced security risks}} \\
\hline

\multirow{5}{0.6cm}[-0.5ex]{\rotatebox{90}{\makecell[c]{\textbf{Technical and}\\[-0.4ex]\textbf{Architectural}\\[-0.4ex]\textbf{Controls}}}}
  & Secure protocols \& encryption    & \makebox[0pt][c]{\(\bullet\)} & \makebox[0pt][c]{\(\bullet\)} & \makebox[0pt][c]{\(\circ\)} & \makebox[0pt][c]{\(\bullet\)} & \makebox[0pt][c]{\(\bullet\)} & & & & & & & & & & & & & & \makebox[0pt][c]{\(\circ\)} & & \makebox[0pt][c]{\(\bullet\)} & \makebox[0pt][c]{\(\circ\)} & & & & & & & & \\
\cline{2-32}
  & Zero trust enforcement            & \makebox[0pt][c]{\(\circ\)} & \makebox[0pt][c]{\(\bullet\)} & & \makebox[0pt][c]{\(\bullet\)} & \makebox[0pt][c]{\(\bullet\)} & & & & & \makebox[0pt][c]{\(\circ\)} & \makebox[0pt][c]{\(\bullet\)} & & \makebox[0pt][c]{\(\circ\)} & \makebox[0pt][c]{\(\bullet\)} & & & & & & & \makebox[0pt][c]{\(\bullet\)} & \makebox[0pt][c]{\(\circ\)} & & & & & & & & \\
\cline{2-32}
  & Hardware-based integrity          & \makebox[0pt][c]{\(\bullet\)} & \makebox[0pt][c]{\(\bullet\)} & \makebox[0pt][c]{\(\bullet\)} & \makebox[0pt][c]{\(\bullet\)} & \makebox[0pt][c]{\(\circ\)} & & & & & & & & & \makebox[0pt][c]{\(\circ\)} & & \makebox[0pt][c]{\(\circ\)} & & & & & \makebox[0pt][c]{\(\bullet\)} & \makebox[0pt][c]{\(\bullet\)} & & & & & & & & \\
\cline{2-32}
  & Intrusion \& anomaly detection    & \makebox[0pt][c]{\(\bullet\)} & \makebox[0pt][c]{\(\bullet\)} & \makebox[0pt][c]{\(\circ\)} & \makebox[0pt][c]{\(\bullet\)} & \makebox[0pt][c]{\(\circ\)} & & & & & & & & & \makebox[0pt][c]{\(\bullet\)} & & & & \makebox[0pt][c]{\(\bullet\)} & \makebox[0pt][c]{\(\bullet\)} & & \makebox[0pt][c]{\(\bullet\)} & \makebox[0pt][c]{\(\bullet\)} & & & & & & & & \\
\cline{2-32}
  & Secure update \& hardening        & \makebox[0pt][c]{\(\circ\)} & \makebox[0pt][c]{\(\circ\)} & \makebox[0pt][c]{\(\bullet\)} & \makebox[0pt][c]{\(\circ\)} & & & & & & & & & & & & & & & & & \makebox[0pt][c]{\(\circ\)} & & & & & & & & & \\
\hline

\multirow{6}{0.65cm}[-1.5ex]{\rotatebox{90}{\makecell[c]{\textbf{Privacy and}\\[-0.4ex]\textbf{Data}\\[-0.4ex]\textbf{Protection}}}}
  & Privacy-enhancing techs    & & & & & \makebox[0pt][c]{\(\bullet\)} & \makebox[0pt][c]{\(\bullet\)} & & \makebox[0pt][c]{\(\bullet\)} & & & \makebox[0pt][c]{\(\circ\)} & & \makebox[0pt][c]{\(\circ\)} & & & & & & & & & & & & \makebox[0pt][c]{\(\bullet\)} & & & & & \\
\cline{2-32}
  & Decentralised data processing     & \makebox[0pt][c]{\(\circ\)} & \makebox[0pt][c]{\(\circ\)} & & \makebox[0pt][c]{\(\circ\)} & \makebox[0pt][c]{\(\bullet\)} & \makebox[0pt][c]{\(\circ\)} & & \makebox[0pt][c]{\(\circ\)} & & & \makebox[0pt][c]{\(\circ\)} & & \makebox[0pt][c]{\(\circ\)} & & & \makebox[0pt][c]{\(\circ\)} & \makebox[0pt][c]{\(\circ\)} & \makebox[0pt][c]{\(\circ\)} & \makebox[0pt][c]{\(\circ\)} & & & & & & \makebox[0pt][c]{\(\bullet\)} & & & & & \\
\cline{2-32}
  & Cryptographic safeguards          & \makebox[0pt][c]{\(\circ\)} & \makebox[0pt][c]{\(\circ\)} & & \makebox[0pt][c]{\(\circ\)} & \makebox[0pt][c]{\(\bullet\)} & \makebox[0pt][c]{\(\bullet\)} & & \makebox[0pt][c]{\(\bullet\)} & & & \makebox[0pt][c]{\(\circ\)} & & & & & & & & & & \makebox[0pt][c]{\(\circ\)} & & & & \makebox[0pt][c]{\(\bullet\)} & & & & & \\
\cline{2-32}
  & Context-aware consent control     & & & & & \makebox[0pt][c]{\(\bullet\)} & \makebox[0pt][c]{\(\circ\)} & \makebox[0pt][c]{\(\bullet\)} & \makebox[0pt][c]{\(\circ\)} & & \makebox[0pt][c]{\(\circ\)} & \makebox[0pt][c]{\(\bullet\)} & \makebox[0pt][c]{\(\circ\)} & \makebox[0pt][c]{\(\bullet\)} & & & & & & & & & & & & \makebox[0pt][c]{\(\bullet\)} & & & & & \\
\cline{2-32}
  & Privacy risk assessment           & & \makebox[0pt][c]{\(\circ\)} & \makebox[0pt][c]{\(\circ\)} & \makebox[0pt][c]{\(\circ\)} & \makebox[0pt][c]{\(\bullet\)} & \makebox[0pt][c]{\(\bullet\)} & \makebox[0pt][c]{\(\bullet\)} & \makebox[0pt][c]{\(\bullet\)} & \makebox[0pt][c]{\(\circ\)} & \makebox[0pt][c]{\(\bullet\)} & \makebox[0pt][c]{\(\bullet\)} & \makebox[0pt][c]{\(\circ\)} & \makebox[0pt][c]{\(\bullet\)} & & & & & & & & & & & & \makebox[0pt][c]{\(\bullet\)} & & & & & \\
\cline{2-32}
  & Data lifecycle management         & & & & \makebox[0pt][c]{\(\circ\)} & \makebox[0pt][c]{\(\bullet\)} & & & \makebox[0pt][c]{\(\circ\)} & & & \makebox[0pt][c]{\(\circ\)} & & & & & & & & \makebox[0pt][c]{\(\circ\)} & & & & & & \makebox[0pt][c]{\(\bullet\)} & & & & & \\
\hline

\multirow{6}{0.65cm}[-1.5ex]{\rotatebox{90}{\makecell[c]{\textbf{AI-Augmented}\\[-0.4ex]\textbf{Security}\\[-0.4ex]\textbf{Intelligence}}}}
  & Predictive threat modelling       & \makebox[0pt][c]{\(\bullet\)} & \makebox[0pt][c]{\(\bullet\)} & \makebox[0pt][c]{\(\circ\)} & \makebox[0pt][c]{\(\bullet\)} & & & & & & & & & & \makebox[0pt][c]{\(\bullet\)} & & & & \makebox[0pt][c]{\(\bullet\)} & \makebox[0pt][c]{\(\circ\)} & & \makebox[0pt][c]{\(\circ\)} & \makebox[0pt][c]{\(\circ\)} & & & & & & & & \\
\cline{2-32}
  & Context-aware access control      & \makebox[0pt][c]{\(\circ\)} & \makebox[0pt][c]{\(\circ\)} & & \makebox[0pt][c]{\(\circ\)} & \makebox[0pt][c]{\(\circ\)} & & & & & & \makebox[0pt][c]{\(\bullet\)} & & & \makebox[0pt][c]{\(\circ\)} & & & & & & & \makebox[0pt][c]{\(\circ\)} & \makebox[0pt][c]{\(\circ\)} & & & & & & & & \\
\cline{2-32}
  & Predictive maintenance            & & & & & & & & & & & & & & \makebox[0pt][c]{\(\bullet\)} & \makebox[0pt][c]{\(\circ\)} & & \makebox[0pt][c]{\(\circ\)} & \makebox[0pt][c]{\(\bullet\)} & \makebox[0pt][c]{\(\circ\)} & & & & & & & & & \makebox[0pt][c]{\(\circ\)} & & \\
\cline{2-32}
  & Explainable AI \& transparency     & & & & & & & & & & & & & & & & & & & & \makebox[0pt][c]{\(\bullet\)} & & \makebox[0pt][c]{\(\circ\)} & \makebox[0pt][c]{\(\bullet\)} & \makebox[0pt][c]{\(\bullet\)} & \makebox[0pt][c]{\(\circ\)} & & \makebox[0pt][c]{\(\bullet\)} & \makebox[0pt][c]{\(\circ\)} & & \\
\cline{2-32}
  & Behavioural anomaly detection      & & & & \makebox[0pt][c]{\(\circ\)} & \makebox[0pt][c]{\(\circ\)} & & & & & \makebox[0pt][c]{\(\circ\)} & \makebox[0pt][c]{\(\circ\)} & & & \makebox[0pt][c]{\(\circ\)} & & & & & & & \makebox[0pt][c]{\(\circ\)} & \makebox[0pt][c]{\(\bullet\)} & & & \makebox[0pt][c]{\(\circ\)} & & & & & \makebox[0pt][c]{\(\circ\)} \\
\cline{2-32}
  & Adaptive security policies        & \makebox[0pt][c]{\(\bullet\)} & \makebox[0pt][c]{\(\circ\)} & & \makebox[0pt][c]{\(\circ\)} & \makebox[0pt][c]{\(\circ\)} & & & & & & \makebox[0pt][c]{\(\circ\)} & \makebox[0pt][c]{\(\circ\)} & \makebox[0pt][c]{\(\circ\)} & \makebox[0pt][c]{\(\bullet\)} & & & & & & & \makebox[0pt][c]{\(\circ\)} & \makebox[0pt][c]{\(\bullet\)} & & & & & & & & \makebox[0pt][c]{\(\circ\)} \\
\hline

\multirow{7}{0.65cm}[-1ex]{\rotatebox{90}{\makecell[c]{\textbf{Governance, Ethics,}\\[-0.4ex]\textbf{and Human}\\[-0.4ex]\textbf{Oversight}}}}
  & Policy-aware design               & & & & & \makebox[0pt][c]{\(\circ\)} & & \makebox[0pt][c]{\(\circ\)} & & \makebox[0pt][c]{\(\circ\)} & \makebox[0pt][c]{\(\bullet\)} & \makebox[0pt][c]{\(\bullet\)} & \makebox[0pt][c]{\(\bullet\)} & \makebox[0pt][c]{\(\bullet\)} & & & & & & & & & & & & \makebox[0pt][c]{\(\circ\)} & & & & & \\
\cline{2-32}
  & Ethical AI guidelines             & & & & & & & & & & & & & & & & & & & & \makebox[0pt][c]{\(\circ\)} & & \makebox[0pt][c]{\(\circ\)} & \makebox[0pt][c]{\(\bullet\)} & \makebox[0pt][c]{\(\circ\)} & \makebox[0pt][c]{\(\bullet\)} & \makebox[0pt][c]{\(\bullet\)} & \makebox[0pt][c]{\(\circ\)} & & & \\
\cline{2-32}
  & Transparency \& auditability      & & & & \makebox[0pt][c]{\(\circ\)} & \makebox[0pt][c]{\(\circ\)} & \makebox[0pt][c]{\(\circ\)} & \makebox[0pt][c]{\(\circ\)} & \makebox[0pt][c]{\(\circ\)} & & \makebox[0pt][c]{\(\bullet\)} & \makebox[0pt][c]{\(\bullet\)} & & & & & & & & & \makebox[0pt][c]{\(\bullet\)} & \makebox[0pt][c]{\(\circ\)} & \makebox[0pt][c]{\(\bullet\)} & \makebox[0pt][c]{\(\bullet\)} & \makebox[0pt][c]{\(\bullet\)} & \makebox[0pt][c]{\(\circ\)} & & \makebox[0pt][c]{\(\bullet\)} & \makebox[0pt][c]{\(\circ\)} & & \\
\cline{2-32}
  & Stakeholder engagement            & & & & & & \makebox[0pt][c]{\(\circ\)} & \makebox[0pt][c]{\(\bullet\)} & & \makebox[0pt][c]{\(\bullet\)} & \makebox[0pt][c]{\(\bullet\)} & \makebox[0pt][c]{\(\bullet\)} & \makebox[0pt][c]{\(\bullet\)} & \makebox[0pt][c]{\(\circ\)} & & & & & & & & & & \makebox[0pt][c]{\(\bullet\)} & \makebox[0pt][c]{\(\circ\)} & \makebox[0pt][c]{\(\circ\)} & \makebox[0pt][c]{\(\circ\)} & \makebox[0pt][c]{\(\circ\)} & \makebox[0pt][c]{\(\circ\)} & & \\
\cline{2-32}
  & Human-in-the-loop oversight       & & & & \makebox[0pt][c]{\(\circ\)} & & & & & & \makebox[0pt][c]{\(\circ\)} & & & & \makebox[0pt][c]{\(\bullet\)} & \makebox[0pt][c]{\(\circ\)} & & & \makebox[0pt][c]{\(\circ\)} & & & & \makebox[0pt][c]{\(\circ\)} & \makebox[0pt][c]{\(\circ\)} & & & & & & & \\
\cline{2-32}
  & Digital ethics                    & & & & & \makebox[0pt][c]{\(\bullet\)} & \makebox[0pt][c]{\(\bullet\)} & \makebox[0pt][c]{\(\bullet\)} & \makebox[0pt][c]{\(\bullet\)} & & \makebox[0pt][c]{\(\bullet\)} & \makebox[0pt][c]{\(\bullet\)} & \makebox[0pt][c]{\(\circ\)} & \makebox[0pt][c]{\(\circ\)} & & & & & & & \makebox[0pt][c]{\(\bullet\)} & & & \makebox[0pt][c]{\(\bullet\)} & \makebox[0pt][c]{\(\bullet\)} & \makebox[0pt][c]{\(\bullet\)} & \makebox[0pt][c]{\(\bullet\)} & \makebox[0pt][c]{\(\circ\)} & & & \\
\cline{2-32}
  & Organisational accountability     & & & & & & & & & & \makebox[0pt][c]{\(\bullet\)} & \makebox[0pt][c]{\(\bullet\)} & & \makebox[0pt][c]{\(\circ\)} & & & & & & & & & & & & & & & & & \\
\hline

\multirow{7}{0.65cm}[-2ex]{\rotatebox{90}{\makecell[c]{\textbf{Resilience}\\[-0.4ex]\textbf{and}\\[-0.4ex]\textbf{Self-Adaptation}}}}
  & Autonomous self-healing           & \makebox[0pt][c]{\(\circ\)} & \makebox[0pt][c]{\(\circ\)} & \makebox[0pt][c]{\(\circ\)} & & & & & & & & & & & \makebox[0pt][c]{\(\bullet\)} & \makebox[0pt][c]{\(\bullet\)} & & \makebox[0pt][c]{\(\bullet\)} & \makebox[0pt][c]{\(\bullet\)} & \makebox[0pt][c]{\(\bullet\)} & & & & & & & & & & & \\
\cline{2-32}
  & Redundancy \& failover            & \makebox[0pt][c]{\(\circ\)} & \makebox[0pt][c]{\(\circ\)} & & & & & & & & & & & & \makebox[0pt][c]{\(\circ\)} & \makebox[0pt][c]{\(\bullet\)} & \makebox[0pt][c]{\(\circ\)} & \makebox[0pt][c]{\(\circ\)} & \makebox[0pt][c]{\(\bullet\)} & \makebox[0pt][c]{\(\circ\)} & & & & & & & & & & & \\
\cline{2-32}
  & Synchronisation \& recovery       & & & & \makebox[0pt][c]{\(\bullet\)} & & & & & & & & & & \makebox[0pt][c]{\(\circ\)} & \makebox[0pt][c]{\(\circ\)} & & \makebox[0pt][c]{\(\bullet\)} & \makebox[0pt][c]{\(\circ\)} & \makebox[0pt][c]{\(\bullet\)} & & & & & & & & & & & \\
\cline{2-32}
  & Real-time fault mitigation        & \makebox[0pt][c]{\(\circ\)} & \makebox[0pt][c]{\(\circ\)} & & \makebox[0pt][c]{\(\circ\)} & & & & & & & & & & \makebox[0pt][c]{\(\bullet\)} & \makebox[0pt][c]{\(\circ\)} & & & \makebox[0pt][c]{\(\bullet\)} & \makebox[0pt][c]{\(\circ\)} & & & & & & & & & & & \makebox[0pt][c]{\(\circ\)}\\
\cline{2-32}
  & Dynamic reconfiguration           & \makebox[0pt][c]{\(\bullet\)} & \makebox[0pt][c]{\(\bullet\)} & \makebox[0pt][c]{\(\circ\)} & \makebox[0pt][c]{\(\circ\)} & & & & & & & & & \makebox[0pt][c]{\(\circ\)} & \makebox[0pt][c]{\(\bullet\)} & \makebox[0pt][c]{\(\circ\)} & \makebox[0pt][c]{\(\circ\)} & \makebox[0pt][c]{\(\circ\)} & \makebox[0pt][c]{\(\bullet\)} & \makebox[0pt][c]{\(\circ\)} & & & & & & & & & & & \\
\cline{2-32}
  & Assessment \& testing  & \makebox[0pt][c]{\(\circ\)} & \makebox[0pt][c]{\(\circ\)} & & \makebox[0pt][c]{\(\circ\)} & & & & & & & & & & \makebox[0pt][c]{\(\bullet\)} & \makebox[0pt][c]{\(\bullet\)} & \makebox[0pt][c]{\(\circ\)} & & \makebox[0pt][c]{\(\bullet\)} & \makebox[0pt][c]{\(\circ\)} & & & & & & & & & & & \\
\cline{2-32}
  & Incident response planning        & \makebox[0pt][c]{\(\circ\)} & \makebox[0pt][c]{\(\circ\)} & \makebox[0pt][c]{\(\circ\)} & \makebox[0pt][c]{\(\circ\)} & & & & & & & & & & \makebox[0pt][c]{\(\bullet\)} & \makebox[0pt][c]{\(\bullet\)} & & \makebox[0pt][c]{\(\bullet\)} & \makebox[0pt][c]{\(\bullet\)} & \makebox[0pt][c]{\(\bullet\)} & & & & & & & & & & & \\
\hline

\multirow{8}{0.65cm}[-2.5ex]{\rotatebox{90}{\makecell[c]{\textbf{Distributed}\\[-0.4ex]\textbf{Trust and}\\[-0.4ex]\textbf{Assurance}}}}
  & Blockchain provenance             & \makebox[0pt][c]{\(\circ\)} & \makebox[0pt][c]{\(\circ\)} & & \makebox[0pt][c]{\(\bullet\)} & & & & & & & \makebox[0pt][c]{\(\circ\)} & & & & & & \makebox[0pt][c]{\(\circ\)} & & \makebox[0pt][c]{\(\circ\)} & & \makebox[0pt][c]{\(\bullet\)} & & & & & & & & & \\
\cline{2-32}
  & Verifiable computation            & & & & \makebox[0pt][c]{\(\circ\)} & & & & & & & & & & & & & & & \makebox[0pt][c]{\(\circ\)} & \makebox[0pt][c]{\(\bullet\)} & & \makebox[0pt][c]{\(\circ\)} & & \makebox[0pt][c]{\(\circ\)} & & & \makebox[0pt][c]{\(\circ\)} & \makebox[0pt][c]{\(\circ\)} & & \\
\cline{2-32}
  & Decentralised identity            & \makebox[0pt][c]{\(\circ\)} & \makebox[0pt][c]{\(\circ\)} & & \makebox[0pt][c]{\(\circ\)} & \makebox[0pt][c]{\(\circ\)} & & & & \makebox[0pt][c]{\(\bullet\)} & & & & & & & & & & & & \makebox[0pt][c]{\(\bullet\)} & \makebox[0pt][c]{\(\circ\)} & & & & & & & & \\
\cline{2-32}
  & Trusted exec. environments    & & \makebox[0pt][c]{\(\bullet\)} & \makebox[0pt][c]{\(\circ\)} & \makebox[0pt][c]{\(\bullet\)} & \makebox[0pt][c]{\(\bullet\)} & & & & & & & & & & & & & & & & \makebox[0pt][c]{\(\circ\)} & \makebox[0pt][c]{\(\bullet\)} & & & & & & & & \\
\cline{2-32}
  & Consensus mechanisms              & & \makebox[0pt][c]{\(\circ\)} & & \makebox[0pt][c]{\(\bullet\)} & & & & & \makebox[0pt][c]{\(\circ\)} & & \makebox[0pt][c]{\(\circ\)} & & & & & & \makebox[0pt][c]{\(\bullet\)} & \makebox[0pt][c]{\(\circ\)} & \makebox[0pt][c]{\(\bullet\)} & & & & & & & & & & & \\
\cline{2-32}
  & Distributed Logging               & & & & \makebox[0pt][c]{\(\bullet\)} & & & & & & & \makebox[0pt][c]{\(\circ\)} & & & \makebox[0pt][c]{\(\circ\)} & & & & & \makebox[0pt][c]{\(\circ\)} & \makebox[0pt][c]{\(\circ\)} & \makebox[0pt][c]{\(\circ\)} & \makebox[0pt][c]{\(\bullet\)} & & \makebox[0pt][c]{\(\circ\)} & & & & & & \\
\cline{2-32}
  & Trust scoring \& reputation      & & \makebox[0pt][c]{\(\circ\)} & \makebox[0pt][c]{\(\circ\)} & \makebox[0pt][c]{\(\circ\)} & & & & & & \makebox[0pt][c]{\(\circ\)} & & & & \makebox[0pt][c]{\(\circ\)} & & & & & & & \makebox[0pt][c]{\(\bullet\)} & \makebox[0pt][c]{\(\circ\)} & & & & & & & & \makebox[0pt][c]{\(\circ\)} \\
\cline{2-32}
  & Token-based access control      & & \makebox[0pt][c]{\(\circ\)} & & \makebox[0pt][c]{\(\circ\)} & \makebox[0pt][c]{\(\circ\)} & & & & & \makebox[0pt][c]{\(\circ\)} & \makebox[0pt][c]{\(\bullet\)} & & & & & & & & & & \makebox[0pt][c]{\(\bullet\)} & \makebox[0pt][c]{\(\circ\)} & & & & & & & & \makebox[0pt][c]{\(\circ\)} \\
\hline

\multirow{5}{0.65cm}[-1.5ex]{\rotatebox{90}{\makecell[c]{\textbf{Human}\\[-0.4ex]\textbf{Trust-}\\[-0.4ex]\textbf{worthiness}}}}
  & Security \& ethics training       & \makebox[0pt][c]{\(\circ\)} & & \makebox[0pt][c]{\(\circ\)} & & \makebox[0pt][c]{\(\circ\)} & & \makebox[0pt][c]{\(\circ\)} & & & \makebox[0pt][c]{\(\circ\)} & & & & & & & & & & & & \makebox[0pt][c]{\(\circ\)} & \makebox[0pt][c]{\(\circ\)} & & \makebox[0pt][c]{\(\circ\)} & \makebox[0pt][c]{\(\bullet\)} & \makebox[0pt][c]{\(\bullet\)} & \makebox[0pt][c]{\(\circ\)} & \makebox[0pt][c]{\(\bullet\)} & \makebox[0pt][c]{\(\bullet\)} \\
\cline{2-32}
  & Continuous capacity building      & \makebox[0pt][c]{\(\circ\)} & & \makebox[0pt][c]{\(\circ\)} & & \makebox[0pt][c]{\(\circ\)} & & \makebox[0pt][c]{\(\circ\)} & & & \makebox[0pt][c]{\(\circ\)} & & & \makebox[0pt][c]{\(\circ\)} & & & & & & & & & \makebox[0pt][c]{\(\circ\)} & \makebox[0pt][c]{\(\circ\)} & & \makebox[0pt][c]{\(\circ\)} & \makebox[0pt][c]{\(\bullet\)} & \makebox[0pt][c]{\(\bullet\)} & \makebox[0pt][c]{\(\circ\)} & \makebox[0pt][c]{\(\bullet\)} & \makebox[0pt][c]{\(\bullet\)} \\
\cline{2-32}
  & Human-centric design              & & & & & \makebox[0pt][c]{\(\circ\)} & & & & & & & & & & & & & & & \makebox[0pt][c]{\(\circ\)} & & & & \makebox[0pt][c]{\(\circ\)} & & & \makebox[0pt][c]{\(\bullet\)} & \makebox[0pt][c]{\(\bullet\)} & \makebox[0pt][c]{\(\circ\)} & \makebox[0pt][c]{\(\bullet\)} \\
\cline{2-32}
  & Usability \& interface design     & & & & & & & & & & & & & & \makebox[0pt][c]{\(\circ\)} & \makebox[0pt][c]{\(\circ\)} & & \makebox[0pt][c]{\(\circ\)} & \makebox[0pt][c]{\(\circ\)} & & \makebox[0pt][c]{\(\circ\)} & & & & \makebox[0pt][c]{\(\circ\)} & & & \makebox[0pt][c]{\(\bullet\)} & \makebox[0pt][c]{\(\bullet\)} & \makebox[0pt][c]{\(\circ\)} & \\
\cline{2-32}
  & Shared responsibility  & & & & & \makebox[0pt][c]{\(\circ\)} & & & & & \makebox[0pt][c]{\(\bullet\)} & \makebox[0pt][c]{\(\circ\)} & & \makebox[0pt][c]{\(\circ\)} & \makebox[0pt][c]{\(\circ\)} & & & & & & & & \makebox[0pt][c]{\(\circ\)} & & & \makebox[0pt][c]{\(\circ\)} & \makebox[0pt][c]{\(\bullet\)} & \makebox[0pt][c]{\(\circ\)} & & \makebox[0pt][c]{\(\circ\)} & \makebox[0pt][c]{\(\bullet\)} \\
\hline

\end{tabular}
\end{table*}

\section{Methodology}
\label{sec:methodology}

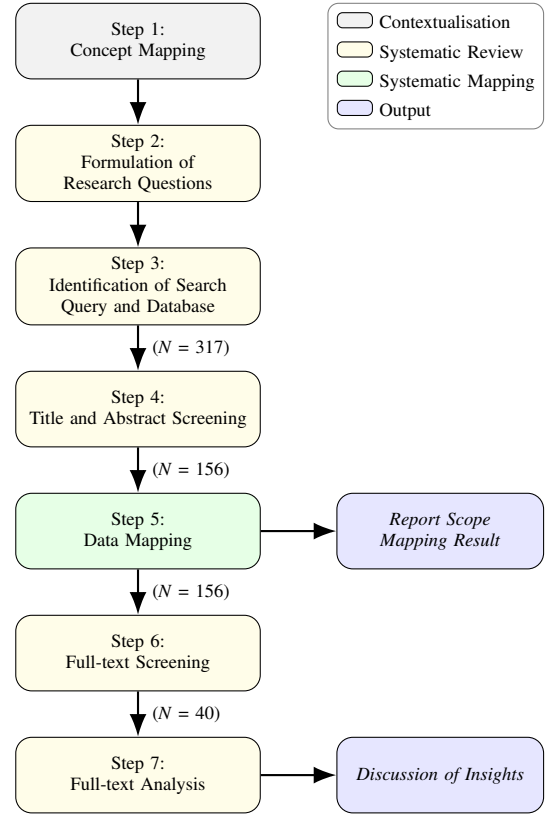
\begin{figure}[ht]
    \centering
    \tikzset{
        process/.style={ rectangle, rounded corners=6pt, draw=black, fill=gray!10, align=center, text width=3cm, minimum height=1cm, font=\scriptsize },
        review/.style={ fill=yellow!10 },
        mapping/.style={ fill=green!10 },
        legend/.style={ rounded corners=2pt },
        output/.style={ fill=blue!10, text width=2.5cm, font=\scriptsize\itshape },
        arrow/.style={ -{Latex[length=3mm]}, thick },
        label/.style={ midway, font=\scriptsize, fill=white, inner sep=1pt, xshift=4pt }
    }

    \begin{tikzpicture}[node distance=0.6cm and 1cm]
        \node[process] (s1) {Step 1:\\Concept Mapping};
        \node[process, review, below=of s1] (s2) {Step 2:\\Formulation of Research Questions};
        \node[process, review, below=of s2] (s3) {Step 3:\\Identification of Search Query and Database};
        \node[process, review, below=of s3] (s4) {Step 4:\\Title and Abstract Screening};
        \node[process, mapping, below=of s4] (s5) {Step 5:\\Data Mapping};
        \node[process, review, below=of s5] (s6) {Step 6:\\Full-text Screening};
        \node[process, review, below=of s6] (s7) {Step 7:\\Full-text Analysis};

        \node[process, output, right=of s5] (r1) {Report Scope Mapping Result};
        \node[process, output, right=of s7] (r2) {Discussion of Insights};

        \draw[arrow] (s1) -- (s2) node[label, right]{};
        \draw[arrow] (s2) -- (s3) node[label, right]{};
        \draw[arrow] (s3) -- (s4) node[label, right]{(\textit{N} = 317)};
        \draw[arrow] (s4) -- (s5) node[label, right]{(\textit{N} = 156)};
        \draw[arrow] (s5) -- (s6) node[label, right]{(\textit{N} = 156)};
        \draw[arrow] (s6) -- (s7) node[label, right]{(\textit{N} = 40)};

        \draw[arrow] (s5) -- (r1);
        \draw[arrow] (s7) -- (r2);

        \node[draw=gray, fill=white, align=left, font=\scriptsize, rounded corners=4pt, anchor=north east] at (current bounding box.north east) (legend) {
            \tikz{\node[process, legend, text width=0.3cm, minimum height=0.2cm]{};} Contextualisation\\[3pt]
            \tikz{\node[process, legend, review, text width=0.3cm, minimum height=0.2cm]{};} Systematic Review\\[3pt]
            \tikz{\node[process, legend, mapping, text width=0.3cm, minimum height=0.2cm]{};}
            Systematic Mapping\\[3pt]
            \tikz{\node[process, output, legend, text width=0.3cm, minimum height=0.2cm]{};} Output
        };
    \end{tikzpicture}

    \caption{Overview of the systematic review process. Steps 2-5 correspond to abstract-based mapping, while Steps 6-7 represent the full-text review and analysis. Outputs from Steps 5 and 7 show the scope mapping results and discussion of insights, respectively.}
    \label{fig:review_steps}
\end{figure}

This systematic literature review follows a structured, step-by-step approach, combining the rigour of a systematic review \cite{kitchenham2015evidence} with the broad overview provided by systematic mapping techniques \cite{Petersen2015guidelines}. In the first phase, a scoping-style mapping was performed using information extracted from paper abstracts to visualise trends and coverage of trust-related issues in DT research. This was followed by a full-text review aimed at identifying specific research gaps and gaining deeper insights into trust considerations. The overall workflow is summarised in Figure~\ref{fig:review_steps}, which illustrates the main steps of the study: concept mapping and contextualisation, formulation of research questions, identification of search queries and databases, abstract screening, data mapping with reporting of scope mapping results, full-text screening, and full-text analysis leading to discussion of insights.

\subsection{Concept Mapping and Contextualisation}

Concept mapping and contextualisation \cite{Booth2016systematic} work was conducted to establish the conceptual landscape of the review. Drawing on dependability, factors and methods that could affect users' trust in socio-technical systems were identified and their applicability to DT was examined, considering its characteristics. These factors were then grouped into seven categories of challenges (see Section~\ref{sec:key-challenges}) and seven categories of strategies (see Section~\ref{sec:strategies}). This mapping served as a conceptual scaffold to guide subsequent systematic review activities, ensuring that the review focused on issues directly related to trust in DT deployment rather than general implementation difficulties.

\subsection{Formulation of Research Questions}

The overarching goal of this review is to understand how trustworthiness-related aspects are discussed, particularly within review and survey papers across application domains. Specifically, the study aims to identify which domains are represented in the literature, the types of trust-related challenges recognised, any proposed solutions or strategies, and whether these challenges and responses are domain-specific or generalisable. These considerations are captured in the following research questions:

\begin{itemize}
    \item RQ1: How are application domains represented in DT review papers, and how frequently is trust discussed in each domain?
    \item RQ2: How are trust challenges and trust-related strategies discussed in DT review papers?
    \item RQ3: How have trust challenges and trust-related strategies changed over time in DT papers?
    \item RQ4: How do trust challenges and trust-related strategies vary across different application domains?
\end{itemize}

\subsection{Identification of Search Query and Database}

The search strategy was designed to ensure a comprehensive and replicable retrieval of relevant papers. The searches were conducted on 3 October 2025 using three major scientific databases, Scopus, IEEE Xplore, and Web of Science (WoS), chosen for their broad and complementary coverage of peer-reviewed literature in computer science and engineering.

The search strings were formulated to identify review papers focused on DTs that discuss challenges, strategies, or trust-related issues. Logical operators (AND, OR) and wildcard characters were used to capture keyword variations across databases. Table~\ref{tab:search_queries} summarises the queries and the number of papers retrieved from each source.

\begin{table*}[htbp]
\centering
\scriptsize
\caption{Search queries and results for each database}
\label{tab:search_queries}
\begin{tabularx}{\textwidth}{l>{\raggedright\arraybackslash\ttfamily}Xc}
\hline
\textbf{Database} & \textbf{Search Query} & \textbf{Results} \\
\hline
\textbf{Scopus} &
( ( TITLE ( digital twin ) AND TITLE ( review ) ) AND ( ABS ( challenge ) OR ABS ( strategy ) OR ABS ( strategies ) OR ABS ( trust ) OR ABS ( security ) OR ( cybersecurity ) OR ABS ( privacy ) OR ABS ( resilience ) OR ABS ( ethic ) OR ABS ( governance ) OR ABS ( human error ) ) ) AND ( LIMIT-TO ( SRCTYPE , "j" ) ) AND ( LIMIT-TO ( DOCTYPE , "re" ) OR LIMIT-TO ( DOCTYPE , "ar" ) ) AND ( LIMIT-TO ( LANGUAGE , "English" ) ) &
309 \\
\textbf{IEEE Xplore} &
( "Document Title":"digital twin*" AND "Document Title":"review" ) AND ( "Abstract":"challenge*" OR "Abstract":"strateg*" OR "Abstract":"*security" OR "Abstract":"privacy" OR "Abstract":"trust*" OR "Abstract":"resilience" OR "Abstract":"ethic*" OR "Abstract":"governance" OR "Abstract":"human error" ) &
24 \\
\textbf{Web of Science} &
( "digital twin*" AND "review" ) ( Title ) AND ( "challenge*" OR "strateg*" OR "trust*" OR "security" OR "cybersecurity" OR "privacy" OR "resilience" OR "ethic*" OR "governance" OR "human error*" ) ( Abstract ); filtered by Review Article or Article; English; All Open Access &
194 \\
\hline
\end{tabularx}
\end{table*}

In total, 527 papers were retrieved. Duplicate detection was carried out using Covidence \cite{Covidence2025}, which automatically removed 208 duplicates, followed by the manual removal of two additional duplicates, leaving 317 unique studies for title and abstract screening.

\subsection{Study Selection Criteria}

The study selection process consisted of two stages: title and abstract screening, followed by full-text screening. Inclusion and exclusion criteria were defined to retain only papers relevant to trust-related aspects of DT technologies.

A paper was included if its abstract discussed challenges or strategies affecting DT trustworthiness, rather than focusing solely on general implementation difficulties or performance optimisation. Papers that mentioned relevant keywords only in the context of domain-specific operational challenges, without addressing trust-related implications of DT deployment, were excluded.

From the initial 317 papers retrieved, 161 were excluded for not addressing user trust or related topics, leaving 156 papers for full-text screening. Of these, 40 papers were examined in full, guaranteeing that each major application domain (see Section~\ref{sec:application_domains}) was represented by at least one eligible study. This in-depth assessment provided a clearer understanding of existing gaps, proposed solutions, and emerging domain-specific trends.

\subsection{Data Mapping}

A systematic scoping and mapping analysis was conducted based on the abstracts of the 156 selected papers. The purpose of this mapping was to visualise the distribution of DT research across application domains and to quantify how frequently trust-related aspects appeared in the abstracts. This provides a high-level overview of how the field frames and prioritises trustworthiness-related concerns.

Data extraction and tagging were performed using Covidence, where each paper was assigned tags corresponding to the seven challenges and seven strategies this study identified. Once all papers were tagged, the mapping was generated by aggregating the number of tags per application domain and per year, providing a quantitative view of research focus and trends.

The resulting mapping is presented in Section~\ref{sec:results}, offering a visual overview of the field and forming the foundation for the subsequent full-text review and more detailed analysis of research gaps.

\subsection{Analysis of Results and Discussion of Insights}

The results from the mapping phase were analysed using descriptive statistics and a variety of visual representations to illustrate research trends and the prominence of trust-related discussions across DT domains. Specifically, the analysis included:

\begin{itemize}
    \item Bar chart showing the number of papers that were relevant or irrelevant in each application domain;
    \item Bar chart quantifying how frequently each trust-related challenge and strategy was mentioned across the mapped studies;
    \item Stacked bar charts depicting the temporal evolution of papers addressing different challenges and strategies over the years;
    \item Pie charts illustrating the distribution of research focus on challenges and strategies within each application domain.
\end{itemize}

These visualisations provide a high-level overview of how trust-related concerns are represented in DT research, highlighting both domain-specific patterns and trends over time. The insights gained from this mapping form the foundation for the subsequent full-text review, which aims to explore research gaps, detailed strategies, and domain-specific implications in greater depth.

\section{Results}
\label{sec:results}

The mapping results reported in this study are primarily based on information extracted from abstracts. While this provides an efficient overview of the research landscape, it may not capture all nuances present in full texts. Some papers may discuss trust-related factors in greater detail than is reflected in their abstracts. Nevertheless, studies that meaningfully engage with trustworthiness are likely to highlight these considerations in the abstract. Accordingly, the mapping effectively reflects the visibility and prioritisation of trust-related issues in the current DT literature, while the subsequent full-text review provides a deeper understanding of research gaps and underlying trends.

\bigskip

\noindent\textbf{RQ1: How are application domains represented in DT review papers, and how frequently is trust discussed in each domain?}

\begin{figure*}[ht]
    \centering
    \includegraphics[width=\linewidth]{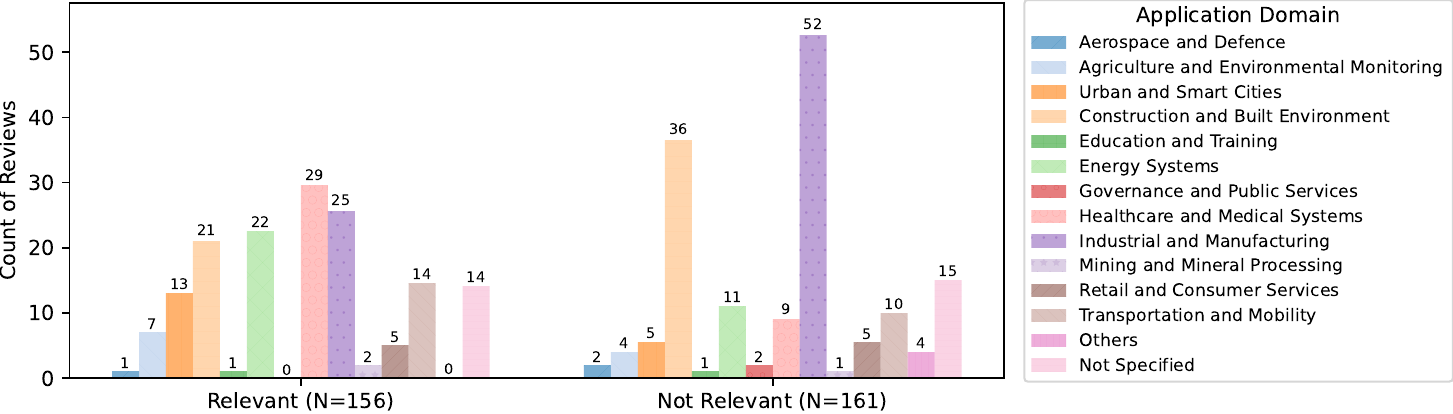}
    \caption{Number of relevant (left) and irrelevant (right) papers per application domain identified during abstract screening.}
    \label{fig:application_domain}
\end{figure*}

Figure~\ref{fig:application_domain} shows the distribution of relevant and irrelevant review papers across different application domains identified during the abstract screening stage. No review papers were found in the Finance and Banking domain under the applied search criteria, and no papers in the Governance and Public Service domain explicitly addressed trustworthiness in the abstract. In contrast, Healthcare, Energy, Urban and Smart Cities domains contain the majority of trust-relevant papers, with more than half of the identified reviews in these areas engaging with trust-related topics. Conversely, in domains such as Construction and Built Environment, and Industrial and Manufacturing Systems, a greater number of papers were classified as irrelevant. Four papers, including those in CPS Testing, Innovation Management, and Software Engineering, do not correspond to the key application domains identified in Section~\ref{sec:application_domains} and were grouped into an "Others" category. While some patterns can be observed in the remaining domains, the small number of identified review papers in these categories limits the representativeness of their proportions.

\bigskip

\noindent\textbf{RQ2: How are trust challenges and trust-related techniques discussed in DT papers?}

\begin{figure*}[ht]
    \centering
    \includegraphics[width=\linewidth]{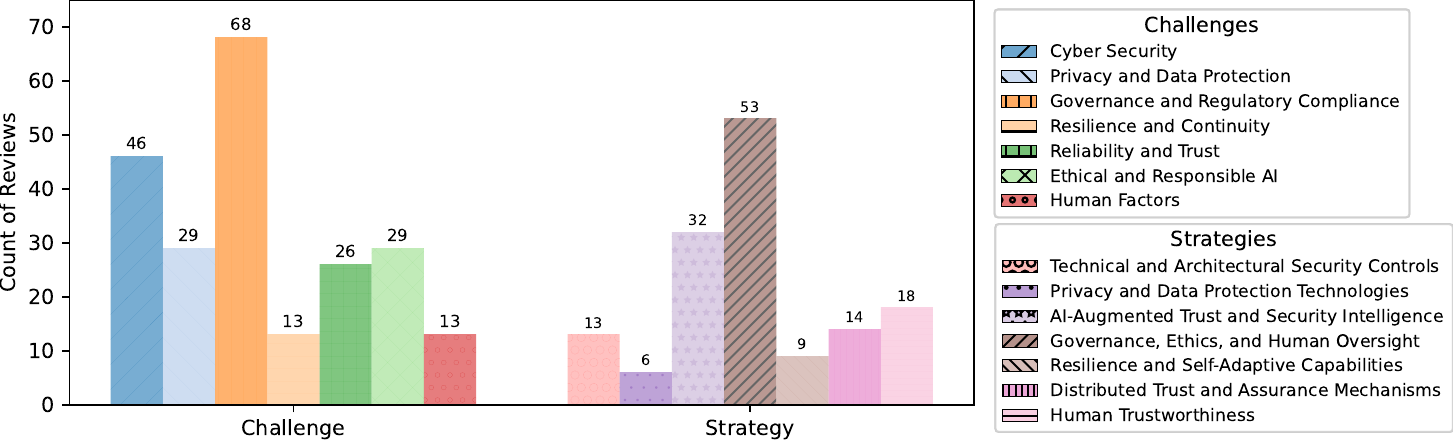}
    \caption{Bar chart showing the frequency of mentions for each trust-related challenge (left) and mitigation strategy (right) across the mapped studies.}
    \label{fig:trustworthiness_focus}
\end{figure*}

Figure~\ref{fig:trustworthiness_focus} shows the frequency of mentions for each trust-related challenge and strategy  across the mapped DT review papers. Among the challenges, Governance and Regulatory Compliance is the most frequently reported, followed by Cyber Security, Privacy and Data Protection, and Ethical and Responsible AI. Less frequently mentioned challenges include Resilience, Human Factors, and Reliability. Regarding strategies, Governance, Ethics, and Human Oversight appears most often, with AI-Augmented Trust and Security Intelligence also reported in a number of studies. Technical measures such as security controls and privacy technologies, as well as strategies related to resilience and distributed trust, are less frequently mentioned.

\bigskip

\noindent\textbf{RQ3: How have trust challenges and trust-related techniques changed over time in DT review papers?}

\begin{figure*}[ht]
    \centering
    \includegraphics[width=\linewidth]{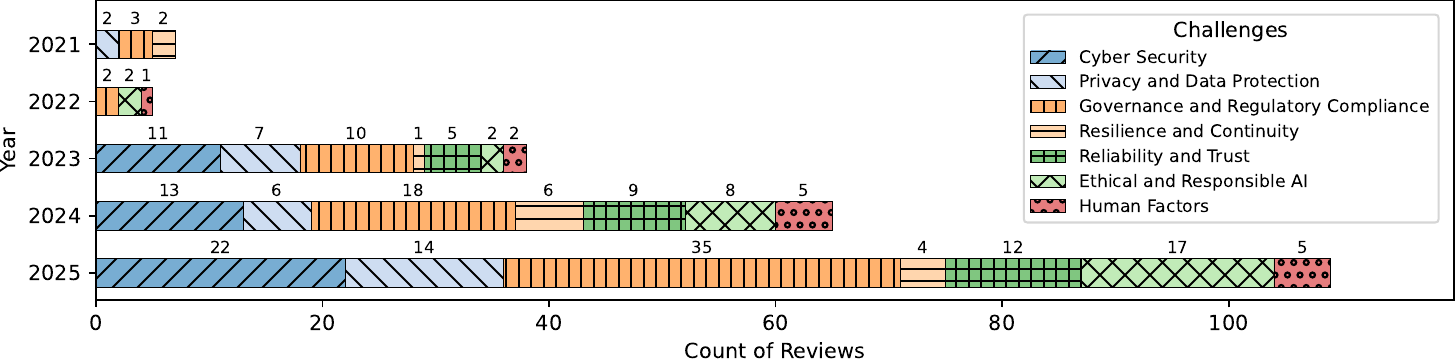}
    \caption{Stacked bar chart illustrating the temporal evolution of papers addressing different challenges over the years.}
    \label{fig:challenges_by_year}
\end{figure*}

Figure~\ref{fig:challenges_by_year} shows the evolution of trust-related challenges discussed in DT review papers from 2021 to 2025. In 2021 and 2022, only a few addressed trust-related challenges, with main focus on Governance and Regulatory Compliance, and isolated mentions of Privacy and Data Protection, Resilience and Continuity, and Ethical AI. From 2023 onwards, there is a marked increase in the number of papers discussing trust issues, with the most frequently addressed challenges being Governance and Regulatory Compliance, and Cyber Security. Other areas such as Reliability and Trust, Ethical AI and Human Factors also begin to appear more consistently. Across these years, the proportion of focus among most challenges remains relatively stable, with the exception of Resilience and Continuity, which remains less frequently addressed compared to other challenges. Overall, the data indicate that the attention of the community towards trust challenges in DT reviews has grown substantially since 2023, with an increasing number of papers considering multiple dimensions of trust.

\begin{figure*}[ht]
    \centering
    \includegraphics[width=\linewidth]{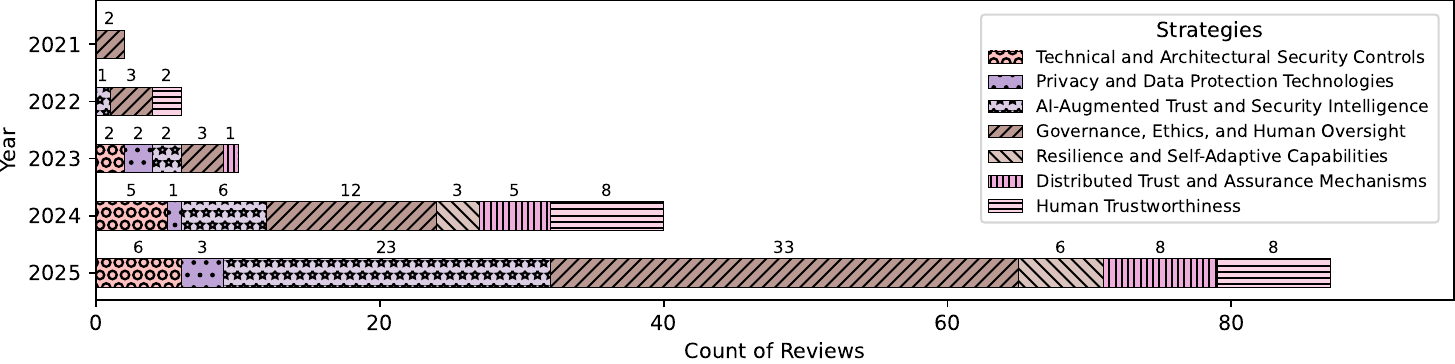}
    \caption{Stacked bar chart illustrating the temporal evolution of papers addressing different mitigation strategies over the years.}
    \label{fig:strategies_by_year}
\end{figure*}

Figure~\ref{fig:strategies_by_year} shows the evolution of trust-related strategies discussed in DT review papers during the same period. Similarly, in 2021 and 2022, very few papers addressed trust strategies, with Governance, Ethics, and Human Oversight appearing most frequently and occasional mentions of AI-Augmented Trust and Human Trustworthiness. From 2023 onwards, there is a marked increase in the number of papers discussing trust strategies. Governance, Ethics, and Human Oversight remains the most frequently addressed strategy, while AI-Augmented Trust and Security Intelligence shows a sharp rise, particularly in 2025. Other strategies, including Technical and Architectural Security Controls, Privacy and Data Protection Technologies, Distributed Trust, Resilience, and Human Trustworthiness, also begin to appear more consistently. Overall, these results indicate that the community has increasingly considered multiple strategies to address trust in DTs since 2023.

\bigskip

\noindent\textbf{RQ4: How do trust challenges and trust-related techniques vary across different application domains?}

\begin{figure*}[ht]
    \centering
    \includegraphics[width=\linewidth]{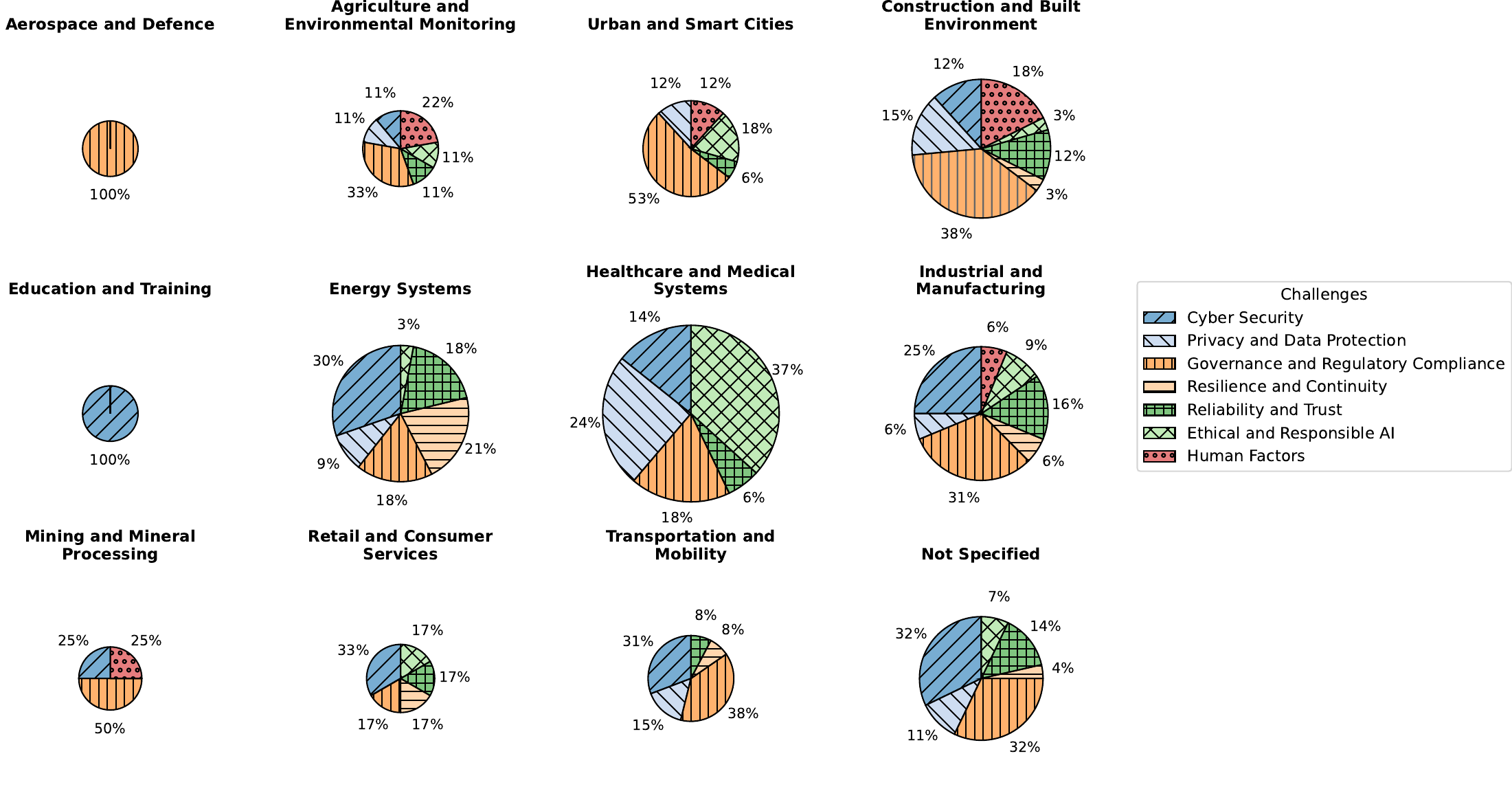}
    \caption{Pie charts illustrating the distribution of research focus on challenges across application domains.}
    \label{fig:challenges_by_ad}
\end{figure*}

Figure~\ref{fig:challenges_by_ad} illustrates the distribution of trust-related challenges across different application domains. Across most domains, Governance and Regulatory Compliance receives the largest proportion of focus, followed by Cyber Security, and Privacy and Data Protection. Exceptions are observed in the Healthcare and Medical domain, where Ethical and Responsible AI, and Privacy and Data Protection are addressed more frequently than Governance or Cyber Security. In the Energy domain, Resilience and Continuity, as well as Reliability and Trust, are frequently discussed challenges after Cyber Security. Other domains, including Aerospace and Defence, Agriculture and Environment Monitoring, Urban and Smart Cities, Construction and Built Environment, Industry and Manufacturing, and Mining and Mineral Processing, highlight Human Factors as a challenge to the trustworthiness of DT systems. In contrast, the Education and Training domain reports only Cyber Security, however, as this is based on a single review paper, no clear trend can be inferred.

\begin{figure*}[ht]
    \centering
    \includegraphics[width=\linewidth]{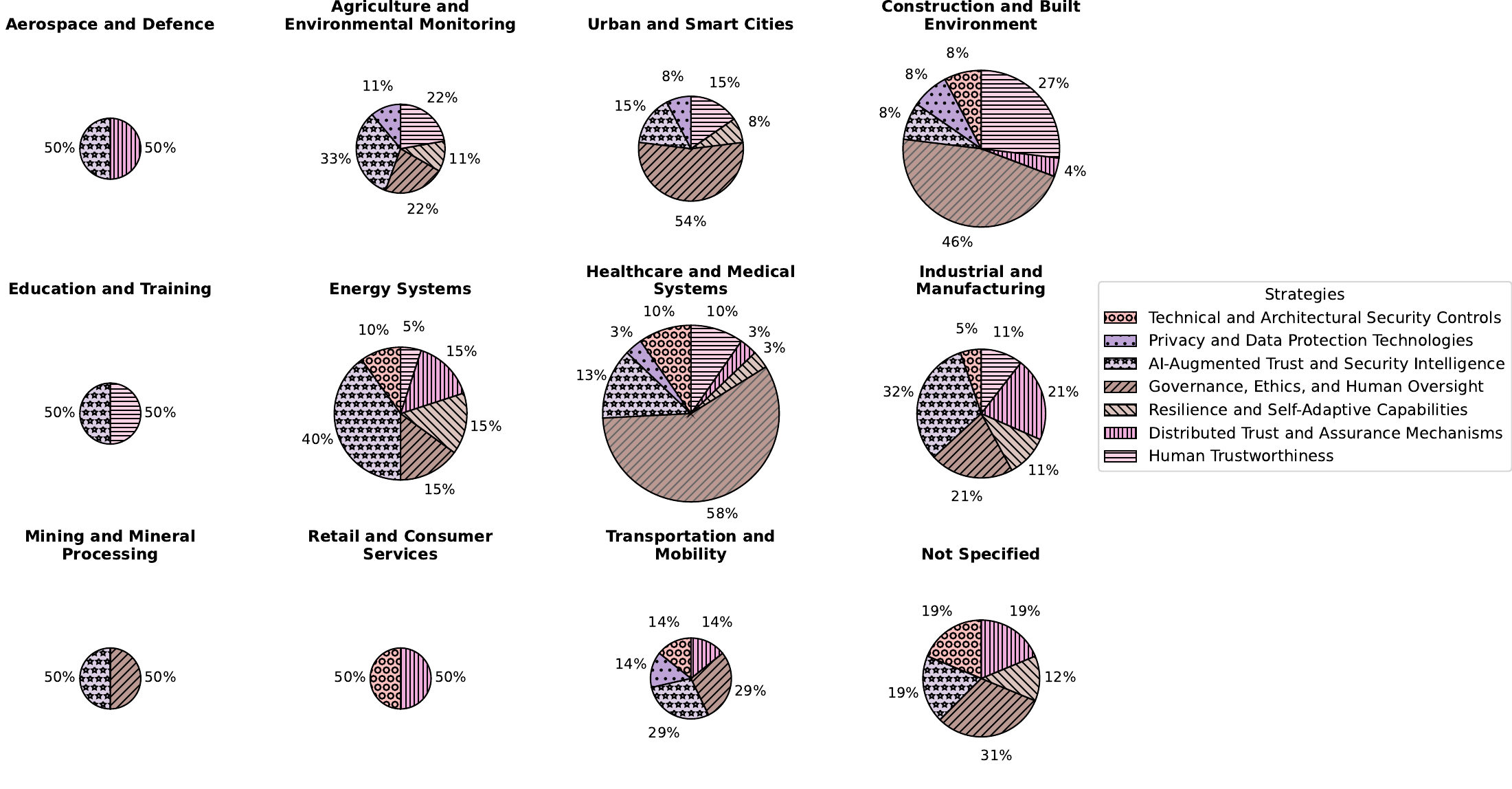}
    \caption{Pie charts illustrating the distribution of research focus on mitigation strategies across application domains.}
    \label{fig:strategies_by_ad}
\end{figure*}

Figure~\ref{fig:strategies_by_ad} shows the distribution of trust-related strategies across application domains. Across most domains, Governance, Ethics and Human Oversight, and AI-Augmented Trust and Security Intelligence receive the greatest attention. Distributed Trust and Assurance Mechanisms are particularly emphasised in the Energy domain, and in Industrial and Manufacturing systems. Strategies aimed at enhancing Human Trustworthiness appear in Agriculture and Environment Monitoring, Urban and Smart Cities, Construction and Built Environment, Healthcare and Medical, and Industrial and Manufacturing systems, closely reflecting the prominence of Human Factors observed among the corresponding trust challenges.

\bigskip

Following the abstract screening, a full-text review on 40 papers was conducted to obtain a more detailed understanding of how trust challenges and related strategies are addressed within each application domain. Table~\ref{tab:fulltext_summary} and Table~\ref{tab:fulltext_summary_continued} summarise the findings from this stage, including the reviewed papers, their specific application use cases, and the trust challenges and strategies discussed. This synthesis provides a more nuanced view of how trustworthiness is conceptualised and operationalised across different DT domains beyond what could be observed from the abstract-level mapping.

\begin{table*}[htbp]
\centering
\scriptsize
\caption{Summary of full-text reviewed papers by application domain}
\label{tab:fulltext_summary}
\begin{tabularx}{\textwidth}{L{3cm} X X}
\hline
\textbf{Domain} & \textbf{Challenges} & \textbf{Strategies} \\
\hline
\begin{minipage}[t]{\linewidth}\raggedright
    \textbf{Industrial and Manufacturing} \\[1em]
    AI \cite{Chen2025aienhanced}; \\ 
    Industry \& Medical \cite{Daraio2025an}; \\
    Industry 5.0 \cite{Domínguez2024digital}; \\
    (I)IoT Security \cite{Mohammed2024systematic}
\end{minipage}
&
\begin{minipage}[t]{\linewidth}\raggedright
    \begin{itemize}[leftmargin=*, topsep=0pt, itemsep=0pt, parsep=0pt, partopsep=0pt]
        \item Inconsistent or low-quality real-time data
        \item Lack of standards \& diverse legacy systems
        \item Absence of recalibration makes twins outdated
        \item Large-scale deployments demand heavy computing
        \item Higher cyber-attack risk from digitalisation
        \item High cost \& strict compliance rules delay adoption
        \item Few large-scale tests create practical uncertainty
        \item Lack of intuitive interfaces \& training causes operator cognitive overload
        \item Lack of trust in AI-driven decisions
        \item Automation disrupt jobs \& create ethical concerns
        \item Insufficient domain experts with AI/DT skills
    \end{itemize}
\end{minipage}
&
\begin{minipage}[t]{\linewidth}\raggedright
    \begin{itemize}[leftmargin=*, topsep=0pt, itemsep=0pt, parsep=0pt, partopsep=0pt]
        \item Cloud-edge architectures, optimized algorithms \& distributed computing for real-time operations
        \item Modular design for scalable, cross-domain operations
        \item Standardised protocols for interoperability
        \item Blockchain \& smart contracts for tamper-proof data
        \item Shared testbeds for innovation \& technology transfer
        \item Blockchain, federated learning \& lightweight encryption for security \& privacy
        \item Continuous monitoring for data \& model reliability
        \item AIoT-driven data management for predictive analytics
        \item Human-centric, immersive \& AI control interfaces
        \item Workforce upskilling \& human-in-the-loop AI
    \end{itemize}
\end{minipage}
\\
\hline

\begin{minipage}[t]{\linewidth}\raggedright
    \textbf{Mining and Mineral Processing} \\[1em]
    Entire Mining Lifecycle \cite{Qu2023digital}
\end{minipage}
&
\begin{minipage}[t]{\linewidth}\raggedright
    \begin{itemize}[leftmargin=*, topsep=0pt, itemsep=0pt, parsep=0pt, partopsep=0pt]
        \item Uncertainties from geological \& process variability
        \item No interoperability standards across systems \& complex organisational structure
        \item Legacy software \& hardware incompatibility
        \item Higher cyber-attack risk from digitalisation
        \item Data-sharing reluctance to protect trade secrets
        \item Skill gaps from insufficient STEM training
    \end{itemize}
\end{minipage}
& 
\begin{minipage}[t]{\linewidth}\raggedright
    \begin{itemize}[leftmargin=*, topsep=0pt, itemsep=0pt, parsep=0pt, partopsep=0pt]
        \item Industry standards \& aligned organisational strategies for interoperability \& security
        \item AI anomaly detection for proactive risk alerts
        \item Real-time analytics for maintenance \& supervision
        \item Remote access \& redundancy for harsh conditions
        \item Immersive interfaces for better user acceptance
        \item STEM reform for DT-related skills
    \end{itemize}
\end{minipage}
\\
\hline

\begin{minipage}[t]{\linewidth}\raggedright
    \textbf{Energy Systems} \\[1em]
    Solar \cite{Kavousifard2024digital}; \\ 
    European Grid \cite{Jørgensen2025digital}; \\
    AI \cite{Abdessadak2025digital}; \\
    Smart Grids \cite{Nabil2024the, Sifat2023towards}
\end{minipage}
&
\begin{minipage}[t]{\linewidth}\raggedright
    \begin{itemize}[leftmargin=*, topsep=0pt, itemsep=0pt, parsep=0pt, partopsep=0pt]
        \item Lack of real-time, high-quality standardised data
        \item Heterogeneous \& legacy technologies hinder adoption
        \item Data volumes create storage \& transmission overhead
        \item Regulatory gap across borders \& legacy regulations
        \item Fragmented data governance and privacy rules
        \item Cyber security, privacy \& data ownership risks
        \item High costs \& misaligned incentives toward physical infrastructure investment
        \item Ethical concerns from opaque AI/ML \& unclear accountability
        \item Environmental impact from energy-intensive AI/ML
    \end{itemize}
\end{minipage}
&
\begin{minipage}[t]{\linewidth}\raggedright
    \begin{itemize}[leftmargin=*, topsep=0pt, itemsep=0pt, parsep=0pt, partopsep=0pt]
        \item Cloud-edge, modular \& HPC for real-time analytics
        \item Open architectures \& standardised protocols
        \item Use Energy Harvesting Wireless Sensors (EHWS)
        \item Establish common data spaces for secure sharing
        \item Blockchain \& encryption for security \& privacy
        \item Robust security frameworks \& compliance (e.g. NIS2)
        \item Create regulatory sandboxes for testing
        \item Data governance \& ethical AI
        \item Academia, industry, policymakers \& government collaboration to overcome technical \& cost barriers
        \item Workforce upskilling \& create DT-specific roles
    \end{itemize}
\end{minipage}
\\
\hline

\begin{minipage}[t]{\linewidth}\raggedright
    \textbf{Construction and Built Environment} \\[1em]
    Energy \cite{Arsecularatne2024review, Sghiri2025leveraging}; \\ 
    Building Development Process \cite{Jahangir2024review}; \\
    Asset Management \cite{Alhadi2025enhancing}; \\
    Construction \cite{Baghdadi2025comprehensive, Hossein2023digital}
\end{minipage}
&
\begin{minipage}[t]{\linewidth}\raggedright
    \begin{itemize}[leftmargin=*, topsep=0pt, itemsep=0pt, parsep=0pt, partopsep=0pt]
        \item Poor data quality \& continuous availability
        \item Heterogeneous protocols, data models \& frameworks
        \item Lack of industry standards \& governance frameworks
        \item Integration with existing systems
        \item Difficulty retrofitting existing or legacy sensors \& buildings
        \item Data volumes create storage \& computation demands
        \item High cost in initial investment (e.g. advanced IT infrastructure) \& workforce training
        \item Cyber security, privacy \& ethical concerns
        \item Unpredictable individual behavioural variability
        \item Fragmented regulatory compliance across jurisdictions
        \item Stakeholders' limited understanding \& acceptance, cultural barriers, \& resistance to digital transformation
        \item Unclear business models \& practical value
        \item Shortage of skilled professionals
    \end{itemize}
\end{minipage}
&
\begin{minipage}[t]{\linewidth}\raggedright
    \begin{itemize}[leftmargin=*, topsep=0pt, itemsep=0pt, parsep=0pt, partopsep=0pt]
        \item HPC, fog, cloud-edge \& microservices for scalability
        \item Open architectures, \& standardised protocols, interfaces \& metadata
        \item Phased implementation \& pilot projects
        \item Cost-effective digitisation \& retrofitting methods for existing assets
        \item Semantic data models \& ontologies for harmonisation
        \item Adaptive control for dynamically occupant behaviour
        \item Robust data governance, monitoring \& validation
        \item Blockchain, encryption, access control \& regular audits for privacy \& data protection
        \item AI anomaly detection, prediction \& optimisation
        \item Middleware platforms for system integration
        \item Develop unified compliance frameworks to address multi‑jurisdiction requirements
        \item Workforce upskilling \& improve client's awareness
    \end{itemize}
\end{minipage}
\\
\hline

\begin{minipage}[t]{\linewidth}\raggedright
    \textbf{Urban and Smart Cities} \\[1em]
    Urban DTs \cite{Lei2023challenges}; \\
    Urban Planning \cite{Shervin2025what}; \\
    Risks \cite{Ariyachandra2023digital}; \\
    Sustainable Cities \cite{Bibri2023synergistic, Weil2023sustainable}
\end{minipage}
&
\begin{minipage}[t]{\linewidth}\raggedright
    \begin{itemize}[leftmargin=*, topsep=0pt, itemsep=0pt, parsep=0pt, partopsep=0pt]
        \item Poor data quality, availability, harmonisation \& synchronisation 
        \item Heterogeneous protocols, data models, frameworks, metadata standards \& definition
        \item Time-varying \& spatial uncertainties
        \item Bandwidth constraints \& reliable sensor coverage
        \item Data volumes create storage \& computation demands
        \item Legacy buildings \& infrastructures hinder adoption
        \item High cost \& software licensing barriers
        \item Cyber security, privacy \& data ownership risks
        \item Lack of validation \& interpretability due to urban complexity
        \item Version drift \& synchronisation between twins
        \item Limited citizen engagement \& opaque governance
        \item Unclear business models \& practical value
        \item Lack of co-creation \& collaboration mechanisms
        \item Fragmented responsibilities across vendors
        \item Socio-technical, cultural \& digital literacy barriers
        \item Shortage of skilled professionals
        \item Environmental impact from energy-intensive models
    \end{itemize}
\end{minipage}
&
\begin{minipage}[t]{\linewidth}\raggedright
    \begin{itemize}[leftmargin=*, topsep=0pt, itemsep=0pt, parsep=0pt, partopsep=0pt]
        \item Define shared purpose, scope \& definitions
        \item Flexible, adaptive development processes \& context-specific design
        \item Open architectures \& standardised protocols
        \item Cloud-edge architectures for real-time operations \& address jurisdictional/compliance constraints
        \item Synthetic data generation to overcome data silos
        \item Establish data sharing networks for inter- organisational data exchange \& improve governance
        \item Federated learning, differential privacy, anonymisation for collaborative analytics \& preserve privacy
        \item Blockchain, encryption, regular audits \& secure cloud services for cyber security
        \item Explainable AI \& human-in-the-loop for transparency
        \item Middleware platforms to retrofit existing buildings
        \item Policy support, regulatory frameworks \& incentives
        \item Engage stakeholders to align objectives \& expected outcomes, \& to justify cost-benefit
        \item Clear responsibilities, accountability \& compliance
        \item Workforce upskilling \& improve public awareness
    \end{itemize}
\end{minipage}
\\
\hline

\end{tabularx}
\end{table*}

\begin{table*}[htbp]
\centering
\caption{Summary of full-text reviewed papers by application domain (continued)}
\scriptsize
\label{tab:fulltext_summary_continued}
\begin{tabularx}{\textwidth}{L{3cm} X X}
\hline
\textbf{Domain} & \textbf{Challenges} & \textbf{Strategies} \\
\hline
\begin{minipage}[t]{\linewidth}\raggedright
    \textbf{Transportation and Mobility} \\[1em]
    Infrastructure Management \cite{Bin2023digital}; \\ 
    Operation \& Maintenance \cite{Sylwia2025digital}; \\
    Railway \cite{Thompson2025revolutionizing}; \\
    EVs \cite{Ali2023review}; \\
    Risks \cite{Vittorio2024risk}
\end{minipage}
&
\begin{minipage}[t]{\linewidth}\raggedright
    \begin{itemize}[leftmargin=*, topsep=0pt, itemsep=0pt, parsep=0pt, partopsep=0pt]
        \item Heterogeneous, high-volume data \& integration
        \item Time-varying \& spatial uncertainties
        \item Lack of standardised domain-specific protocols
        \item Reply on stable connectivity in dynamic environments \& large-scale deployments
        \item High cost \& low effectiveness in minor operations
        \item Cyber security, privacy risks due to  data exchange
        \item Predictions reliability in critical scenarios
        \item Skill gaps \& job security concerns resist adoption
    \end{itemize}
\end{minipage}
&
\begin{minipage}[t]{\linewidth}\raggedright
    \begin{itemize}[leftmargin=*, topsep=0pt, itemsep=0pt, parsep=0pt, partopsep=0pt]
        \item Cloud-edge architectures for real-time operations
        \item Standardised protocols for interoperability
        \item Alternative, hybrid \& resilient network connectivity
        \item Continuous model validation using real-world data
        \item AI/ML predictive analytics for anomaly detection \& uncertainty management
        \item Federated learning, blockchain, routine audits \& robust data governance
        \item Workforce training \& remote support
    \end{itemize}
\end{minipage}
\\
\hline

\begin{minipage}[t]{\linewidth}\raggedright
    \textbf{Aerospace and Defence} \\[1em]
    UAV \& Energy Management Strategy (EMS) \cite{Cyrille2025critical}
\end{minipage}
&
\begin{minipage}[t]{\linewidth}\raggedright
    \begin{itemize}[leftmargin=*, topsep=0pt, itemsep=0pt, parsep=0pt, partopsep=0pt]
        \item Unpredictable flight conditions hinder real-time EMS
        \item No standards for UAV-DT interoperability
        \item Costly, data-heavy EMS limits real-world testing
        \item Cyber risks in UAV-DT communication
    \end{itemize}
\end{minipage}
&
\begin{minipage}[t]{\linewidth}\raggedright
    \begin{itemize}[leftmargin=*, topsep=0pt, itemsep=0pt, parsep=0pt, partopsep=0pt]
        \item 5G/6G ensure reliable \& low-latency data
        \item AI-driven real-time decision \& adaptive control 
        \item AI/ML enable continuous performance learning
        \item Predictive maintenance for safety \& uptime
        \item Blockchain ensures secure, trusted data exchange
    \end{itemize}
\end{minipage}
\\
\hline

\begin{minipage}[t]{\linewidth}\raggedright
    \textbf{Healthcare and Medical Systems} \\[1em]
    Human DTs \cite{Wolfgang2024ethica, Gaffinet2025human}; \\ 
    Intrusion Detection \cite{Thomas2025intrusion}; \\
    Precision Medicine \cite{Armeni2022digital}; \\
    AI \cite{Gabriele2025advances, Cardenas2025technological}
\end{minipage}
&
\begin{minipage}[t]{\linewidth}\raggedright
    \begin{itemize}[leftmargin=*, topsep=0pt, itemsep=0pt, parsep=0pt, partopsep=0pt]
        \item Heterogeneous, noisy real-world data \& integration
        \item Impractical continuous physiological/cognitive data capturing (e.g. invasive methods)
        \item IoMT devices demand careful calibration
        \item Human mobility \& varying environments hinder reliable data transfer
        \item Data volumes create storage \& computation demands
        \item Transmission of sensitive personal/medical data introduces security, privacy \& ethical risks
        \item Complex dynamic multi-level human model (physiological, psychological, socio-economic)
        \item Lack of clinical validation
        \item Model bias \& fairness (racial, gender, demographic)
        \item Low explainability of AI/ML decisions
        \item Lack of intuitive interfaces raises misinterpretation
        \item Lack of adaptive, context-specific guidance
        \item Unequal access to DT infrastructure
        \item Over-reliance on AI leads to mistrust in professionals 
        \item Clinician job security concerns
    \end{itemize}
\end{minipage}
&
\begin{minipage}[t]{\linewidth}\raggedright
    \begin{itemize}[leftmargin=*, topsep=0pt, itemsep=0pt, parsep=0pt, partopsep=0pt]
        \item Cloud-edge, modular frameworks \& URLLC networks for real-time monitoring \& operations
        \item Standardised protocols \& frameworks for interoperability
        \item Data quality pipeline for normalisation
        \item Regulatory compliance frameworks \& centralised data governance for auditability \& data ownership risks
        \item Federated learning, blockchain, encryption, threat modelling \& context-aware IDS for security \& privacy
        \item Robust \& transparent hybrid AI models
        \item Progressive validation \& iterative refinement
        \item Bias auditing, equity safeguards, ethical guidelines, consent mechanisms \& inclusive access
        \item Fill bias gap by GenAI synthetic data
        \item Human centred principles \& user friendly interfaces
        \item Stakeholder involvement throughout design process
        \item Balanced integration with clinical judgment (HDTs as complements)
        \item Workforce upskilling \& educate citizens
    \end{itemize}
\end{minipage}
\\
\hline

\begin{minipage}[t]{\linewidth}\raggedright
    \textbf{Agriculture and Environmental Monitoring} \\[1em]
    Crop \cite{Zhang2025comprehensive}; \\ 
    Dairy \cite{Rao2025computational}; \\
    Livestock \cite{Mun2024systematic}; \\
    Precision Agriculture \cite{Awais2025advancing}
\end{minipage}
&
\begin{minipage}[t]{\linewidth}\raggedright
    \begin{itemize}[leftmargin=*, topsep=0pt, itemsep=0pt, parsep=0pt, partopsep=0pt]
        \item Heterogeneous, high-volume data \& integration
        \item Poor connectivity, sensor reliability \& rural infrastructure
        \item High costs for sensors, infrastructure \& operations, especially for smallholders
        \item Cyber security, privacy \& data ownership risks
        \item Low explainability \& generalisability of AI/ML
        \item Environmental, ethical, human-animal relation \& social concerns
        \item Scalability \& uncertain return on investment
        \item Limited technical skills \& training among users
    \end{itemize}
\end{minipage}
&
\begin{minipage}[t]{\linewidth}\raggedright
    \begin{itemize}[leftmargin=*, topsep=0pt, itemsep=0pt, parsep=0pt, partopsep=0pt]
        \item Cloud-edge architectures for real-time operations
        \item Standardised protocols, semantic integration \& modular frameworks enable interoperability
        \item Federated learning, blockchain \& robust data governance
        \item Advanced explainable AI/ML
        \item Ethical design \& renewable energy
        \item 3D modelling, simulation \& integration with foundation models
        \item Workforce training, capacity building \& user-friendly interfaces
    \end{itemize}
\end{minipage}
\\
\hline

\begin{minipage}[t]{\linewidth}\raggedright
    \textbf{Education and Training} \\[1em]
    Virtual Learning Environments (VLEs) \cite{Kumar2025systematic}
\end{minipage}
&
\begin{minipage}[t]{\linewidth}\raggedright
    \begin{itemize}[leftmargin=*, topsep=0pt, itemsep=0pt, parsep=0pt, partopsep=0pt]
        \item High setup \& AR/VR equipment costs
        \item Limited access from licensing \& hardware needs
        \item Scalability limits in resource-constrained sites
        \item Privacy concerns over student data handling
        \item Complex tools create steep learning curves
        \item Significant instructor training \& design workload
        \item Limited resources hinder balanced participation
    \end{itemize}
\end{minipage}
&
\begin{minipage}[t]{\linewidth}\raggedright
    \begin{itemize}[leftmargin=*, topsep=0pt, itemsep=0pt, parsep=0pt, partopsep=0pt]
        \item Cloud, edge, modular frameworks \& open-source tools cut costs, provide scalable, flexible access
        \item Standardised data for software interoperability
        \item Industry collaboration expands resources
        \item Training strengthens DT teaching capacity
        \item Onboarding support \& progressive learning improves student adoption
    \end{itemize}
\end{minipage}
\\
\hline

\begin{minipage}[t]{\linewidth}\raggedright
    \textbf{Retail and Consumer Services} \\[1em]
    Supply Chains \cite{Le2024digital, Zaidi2024unlocking}
\end{minipage}
&
\begin{minipage}[t]{\linewidth}\raggedright
    \begin{itemize}[leftmargin=*, topsep=0pt, itemsep=0pt, parsep=0pt, partopsep=0pt]
        \item Lack of real-time, high-quality standardised data
        \item High computational demands \& legacy systems
        \item Cyber security, privacy \& trade-secret risks
        \item Low explainability of AI/ML decisions
        \item Misaligned stakeholder goals \& responsibilities
        \item Shortages of digital skills \& strong resistance to organisational digital transformation
    \end{itemize}
\end{minipage}
&
\begin{minipage}[t]{\linewidth}\raggedright
    \begin{itemize}[leftmargin=*, topsep=0pt, itemsep=0pt, parsep=0pt, partopsep=0pt]
        \item Cloud-edge architectures for real-time operations
        \item International standards to ensure interoperability
        \item Blockchain \& data governance for secure sharing
        \item AI/ML dynamic, predictive, explainable analytics
        \item Sustainability \& risk mitigation design
        \item Collaborative DT to align diverse stakeholders
        \item Strengthen workforce digital capabilities
    \end{itemize}
\end{minipage}
\\
\hline

\end{tabularx}
\end{table*}

\section{Discussion}
\label{sec:discussions}

Trust-related considerations in DT research vary significantly across domains, reflecting differences in technological maturity, operational constraints, stakeholder expectations, and societal impact. As DTs transition from experimental prototypes to real-world deployments, reliability, safety, and user acceptance become increasingly critical, making trustworthiness a central concern for both researchers and practitioners.

Our review reveals that discussions of trustworthiness are most prominent in domains where DTs interact directly with diverse user groups or influence decisions with broad societal consequences. Human-centred domains, such as healthcare, energy systems, and smart cities, emphasise transparency, safety, ethical AI practices, and socio-technical implications. In contrast, safety-critical or highly regulated domains, including aerospace and defence, often treat trust implicitly as a property of rigorous engineering, where system integrity, compliance, and reliability are assumed baseline expectations. Technologically-driven domains, such as industrial manufacturing and finance, focus primarily on technical performance, automation, and data integrity, with trustworthiness embedded in system robustness and operational reliability rather than explicitly framed around human-centred concerns. Finally, context-specific domains, such as agriculture, mining, and education, highlight practical alignment with local workflows, resource constraints, and environmental conditions, where trust is closely linked to usability, adaptability, and operational feasibility.

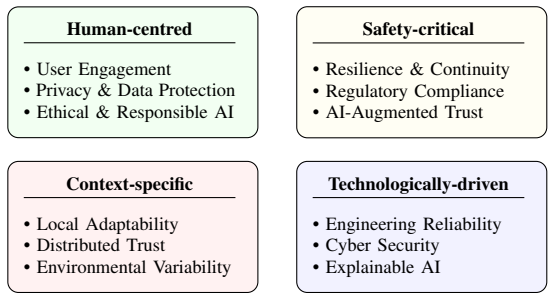
\begin{figure}[ht]
\centering
\begin{tikzpicture}[
    box/.style={
        rectangle, draw, rounded corners, minimum height=1.5cm, 
        align=center, font=\scriptsize, inner sep=6pt
    },
    hc/.style={fill=green!5},
    sc/.style={fill=yellow!5},
    cs/.style={fill=red!5},
    td/.style={fill=blue!5}
]

\node[box, hc] (hc) {
    \begin{minipage}[t]{2.8cm}
        \centering \textbf{Human-centred} \\
        \vspace{-5pt}
        \rule{\linewidth}{0.4pt}
        \vspace{-10pt}
        \begin{flushleft}
            • User Engagement\\
            • Privacy \& Data Protection \\
            • Ethical \& Responsible AI
        \end{flushleft}
    \end{minipage}
};
\node[box, sc, right=0.5cm of hc] (sc) {
    \begin{minipage}[t]{2.8cm}
        \centering \textbf{Safety-critical} \\
        \vspace{-5pt}
        \rule{\linewidth}{0.4pt}
        \vspace{-10pt}
        \begin{flushleft}
            • Resilience \& Continuity \\
            • Regulatory Compliance \\
            • AI-Augmented Trust
        \end{flushleft}
    \end{minipage}
};
\node[box, cs, below=0.3cm of hc] (cs) {
    \begin{minipage}[t]{2.8cm}
        \centering \textbf{Context-specific} \\
        \vspace{-5pt}
        \rule{\linewidth}{0.4pt}
        \vspace{-10pt}
        \begin{flushleft}
            • Local Adaptability \\
            • Distributed Trust \\
            • Environmental Variability
        \end{flushleft}
    \end{minipage}
};
\node[box, td, right=0.5cm of cs] (td) {
    \begin{minipage}[t]{2.8cm}
        \centering \textbf{Technologically-driven} \\
        \vspace{-5pt}
        \rule{\linewidth}{0.4pt}
        \vspace{-10pt}
        \begin{flushleft}
            • Engineering Reliability \\
            • Cyber Security \\
            • Explainable AI
        \end{flushleft}
    \end{minipage}
};

\end{tikzpicture}
\caption{Overview of DT integration categories with major trust and design focus areas.}
\label{fig:dt_integrations}
\end{figure}

From these patterns, we identify four major DT integration categories, including human-centred, safety-critical, context-specific, and technologically-driven, which capture the diversity of trust considerations across application domains (Figure~\ref{fig:dt_integrations}). While certain challenges, such as data heterogeneity, network reliability, and cloud-edge dependencies, are common across all categories, the emphasis on trust-related factors varies considerably. The following subsections examine each category in detail, highlighting domain-specific priorities, strategies, and implications for designing trustworthy DT systems.

\subsection{Human-centred DT Integration}

Human-centred DTs focus on systems that directly interact with or affect individuals and communities, where trustworthiness concerns are closely tied to user experience, ethical considerations, and socio-technical dynamics. Across domains such as healthcare, energy systems, urban management, and transportation, common priorities include ensuring appropriate sensor coverage and robust data pipelines to collect high-quality, heterogeneous data while safeguarding privacy and meeting ethical standards \cite{Cardenas2025technological}. Systems are designed with clear accountability, enabling feedback loops and decision traceability to allow outcomes to be linked back to their sources \cite{Gaffinet2025human}. Data ownership and protection across complex, distributed infrastructures are critical, particularly given the sensitivity of health, mobility, or personal information, and can be supported through secure data-sharing networks that enable standardised, privacy-preserving information exchange across stakeholders \cite{Lei2023challenges}.

Stakeholder and user engagement is central, involving not only end-users but also cross-sector collaboration among academia, industry, and policymakers to align incentives, justify high implementation costs, and establish viable business models \cite{Lei2023challenges}. Human-centred DTs also face challenges related to fairness and bias mitigation, requiring transparent algorithms, inclusive access, and ongoing validation. Responsible AI practices are particularly critical in this category, ensuring that automated decisions remain ethical, accountable, and aligned with the needs and rights of affected individuals and communities \cite{Shervin2025what}. The scale and complexity of these systems often necessitate high-performance computing resources, which introduces trade-offs between computational energy consumption and environmental sustainability. Solutions such as Energy Harvesting Wireless Sensors (EHWS) can help mitigate this tension by supporting energy-efficient sensing and enabling lightweight computation at the edge \cite{Kavousifard2024digital}. Human-machine interfaces are designed to be intuitive, enabling users to understand, interact with, and trust the DT \cite{Armeni2022digital}. Given that these systems often affect large and heterogeneous populations, it is also essential to provide training for staff and education for the general public to raise awareness of the DT's functions, limitations, and societal importance \cite{Ariyachandra2023digital}. Finally, modelling large-scale human and socio-technical systems demands careful prioritisation to capture the most relevant behaviours and interactions while balancing cost, feasibility, and real-world effectiveness.

Overall, human-centred DT integration emphasises a holistic approach that combines technical robustness with ethical, social, and operational considerations, reflecting the need to build systems that are not only accurate but also accountable, effective, and usable by diverse stakeholders.

\subsection{Safety-critical DT Integration}

Safety-critical DTs are deployed in domains where errors or failures can have severe consequences, and trustworthiness is largely treated as an inherent property of system integrity, safety, and regulatory compliance. Domains such as aerospace, defence, energy, and transportation share a strong emphasis on continuous validation, system synchronisation, and reliable data flow to ensure that operational decisions are correct and safe. Given the high stakes involved, low-latency, high-reliability data transmission is essential, and real-world actions are often subject to higher human authorisation, particularly for missions or operational decisions that cannot be risked in live deployment. Regulatory sandboxes play an important role in these contexts, providing risk-free, controlled environments to visualise, test, and validate DT outputs before deployment \cite{Jørgensen2025digital}. These systems must also comply with rigorous regulatory and certification cycles, which govern how DT models are validated, updated, and approved for operational use.

Safety-critical DTs also face complex uncertainties, including temporal and spatial variability, equipment or environmental disturbances, and human behavioural factors. Predictive analytics, AI/ML-driven anomaly detection, and adaptive control are commonly applied to manage these high-dimensional uncertainties and support reliable decision-making \cite{Thompson2025revolutionizing}. Operational resilience is a central requirement, where systems are typically designed with redundancy, failover mechanisms, and fault-tolerant architectures to maintain functionality during disruptions, ensuring that essential operations continue even under degraded conditions \cite{Jørgensen2025digital}. Similarly, synchronised operation across distributed systems requires robust cloud-edge architectures and resilient communication networks to maintain system coherence. Given the complexity and criticality of these environments, expertise from trained DT professionals is essential, both for system development and to provide ongoing monitoring and support when needed \cite{Jørgensen2025digital}.

Finally, technical solutions such as blockchain, federated learning, and secure data-sharing or middleware frameworks contribute to trustworthy operations by ensuring data integrity, traceability, auditability, and secure collaboration across stakeholders. These approaches, while particularly critical in safety-sensitive contexts, also highlight strategies that can inform trustworthiness design in other DT integration domains, bridging lessons from highly regulated, high-risk applications to broader socio-technical systems.

\subsection{Context-specific DT Integration}

Context-specific DTs are characterised by the fact that their trust requirements are shaped primarily by local constraints, decentralised operations, and highly variable physical or organisational environments. They operate in domains where modelling needs, environmental conditions, and organisational practices strongly influence how trustworthiness is understood and maintained. Rather than focusing solely on broad socio-technical issues or stringent safety guarantees, context-specific DTs emphasise practical alignment with local workflows, operational constraints, and situational decision requirements. Application areas span domains such as agriculture, construction, mining, and education, each characterised by heterogeneous settings and varied user expertise.

A major challenge across these domains, such as mining and agriculture, is the need to function within resource- and infrastructure-limited environments, where connectivity, sensor reliability, power availability, or local data scarcity may be inconsistent. Trustworthy performance therefore depends on adaptable architectures, such as cloud-edge balancing, modular components, and energy-efficient hardware, that can accommodate intermittent data availability and dynamic field conditions \cite{Rao2025computational}. Maintaining data consistency across fragmented organisational systems is equally important, as many domains rely on decentralised operations, legacy technologies, or siloed data flows. In some cases, organisations are reluctant to share data because of commercial sensitivities or the need to protect trade secrets, which further limits interoperability and collaboration \cite{Qu2023digital}.

Human and organisational factors play a significant role in establishing trust. Users often have diverse levels of digital literacy and domain knowledge, making clear interfaces, targeted training, and workflow-sensitive system design essential. In many cases, immersive interfaces help practitioners interpret DT outputs, understand model reasoning, and build confidence in DT-supported decisions, particularly when the physical environment is complex or safety-sensitive \cite{Qu2023digital}.

Ethical considerations are also context-dependent and may involve issues related to the welfare of animals, environmental stewardship, or responsible use of learner data in educational settings \cite{Rao2025computational, Kumar2025systematic}. These concerns require governance practices that align with local norms, regulatory expectations, and domain-specific values, ensuring that the deployment of DTs does not unintentionally harm ecosystems, individuals, or communities.

Overall, context-specific DT integration highlights the need for systems that balance technical robustness with practical adaptability, ethical sensitivity, and user-centred design. Trustworthiness emerges from building DTs that are reliable and meaningful within local operating conditions while still aligning with broader principles of transparency, accountability, and responsible data use.

\subsection{Technologically-driven DT Integration}

Technologically-driven DTs are found in domains such as industrial manufacturing, finance, and retail services, where the primary emphasis is on automation, performance, and technical accuracy. In these settings, trustworthiness is often treated as an implicit outcome of engineering reliability rather than a broad socio-technical concern. High-quality, real-time data streams are essential, as models must operate continuously and update rapidly to reflect changing system conditions. This also requires robust data pipelines, continuous monitoring, and recalibration to prevent model drift and ensure that DT outputs remain accurate and operationally relevant.

Cyber security and protection of sensitive information are major challenges, especially given the prevalence of trade secrets, proprietary processes, and market-critical data. Technologies such as blockchain, smart contracts, and secure data-governance frameworks provide tamper-proof records and controlled data access, helping organisations balance collaboration with the need to safeguard competitive advantage \cite{Mohammed2024systematic}. AI/ML capabilities, such as manufacturing optimisation, supply chain prediction, and automated decision systems, must also be explainable to support operator trust, reduce risk, and enable accountability in automated environments \cite{Le2024digital}.

Operational usability remains another core concern. These domains frequently rely on complex, tightly coupled systems, where operators must interpret DT outputs under high time pressure. Intuitive, human-centred interfaces and immersive visualisations help reduce cognitive overload and support correct decision-making \cite{Domínguez2024digital}, while dedicated training is required to develop the advanced digital skills expected in Industry 4.0 and 5.0 environments.

Overall, technologically driven DTs highlight a trust paradigm rooted in engineering performance, security, and system robustness. While less socio-technical than human-centred applications, these domains underscore the need for explainable AI, secure data exchange, resilient architectures, and operator-oriented system design that are increasingly relevant across the broader DT landscape.

\section{Future Works and Research Horizons}
\label{sec:future_work}

As DT technologies continue to evolve and diversify across domains, new design challenges and uncertainties are emerging. Ensuring trustworthy, scalable, and context-appropriate DTs will require ongoing research, methodological development, and collaboration across disciplines. To support this progression, we outline several key avenues for future investigation: establishing trust-by-design practices that guide DT development from the outset, expanding modelling standards to incorporate trust-relevant metadata for platform ecosystems, and examining how architectural choices shape the reliability and governance of DT deployments. These directions highlight areas where current approaches remain fragmented and where sustained research is needed to ensure DTs mature into robust and trustworthy systems.

\subsection{DT Trust-by-Design Framework}

Trust DTs is inherently socio-technical. It reflects not only how well a model performs but also how expectations, risks, responsibilities, and values are distributed among the people who interact with the system. Trust gaps often emerge from conflicting stakeholder priorities, uncertainty about the DT's role, or tensions between performance, transparency, and safety. Therefore, trustworthiness should be embedded from the very earliest stages of design, rather than addressed solely through later assurance activities \cite{Paja2013trust}.

Building the Trust-by-Design framework is not to prescribe universal rules but to provide a structured set of considerations that guide designers, engineers, and domain experts in reasoning about trust implications during the design phase, when foundational decisions about data, modelling, architecture, and governance are made.

Stakeholder perspectives strongly influence which aspects of trust matter most. To capture these perspectives, we interpret DT use through four integration types, which characterise how a DT will be used in practice. Each integration type emphasises different trust considerations, allowing designers to prioritise specific principles early in the design process.

\begin{figure}[ht]
\centering
\begin{tikzpicture}[
    node distance=0.4cm,
    box/.style={rectangle, draw, rounded corners, align=center, minimum width=2.6cm, minimum height=0.6cm, font=\scriptsize},
    decision/.style={diamond, draw, aspect=3, align=center, minimum width=3.5cm, inner sep=2pt, font=\scriptsize},
    arrow/.style={ -{Latex[length=3mm]}, thick },
    label/.style={ midway, font=\scriptsize, fill=white, inner sep=1pt, xshift=-4pt, yshift=2pt },
    labelright/.style={ xshift=8pt },
    hc/.style={fill=green!5},
    sc/.style={fill=yellow!5},
    cs/.style={fill=red!5},
    td/.style={fill=blue!5}
]

\node[decision] (q1) {Failures cause\\serious harm?};
\node[box, sc, right=1cm of q1] (sc) {Safety-critical};
\node[decision, below=0.7cm of q1] (q2) {Direct human\\impact?};
\node[box, hc, right=1cm of q2] (hc) {Human-centred};
\node[decision, below=0.7cm of q2] (q3) {Shaped by\\local constraints?};
\node[box, cs, right=1cm of q3] (cs) {Context-specific};
\node[box, td,  below=0.7cm of q3] (td) {Technologically-driven};

\draw[arrow] (q1) -- (sc) node[label, above]{Yes};
\draw[arrow] (q1) -- (q2) node[label, labelright, right]{No};
\draw[arrow] (q2) -- (hc) node[label, above]{Yes};
\draw[arrow] (q2) -- (q3) node[label, labelright, right]{No};
\draw[arrow] (q3) -- (cs) node[label, above]{Yes};
\draw[arrow] (q3) -- (td) node[label, labelright, right]{No};

\end{tikzpicture}
\caption{Decision flow for identifying DT integration category.}
\label{fig:integration_decision}
\end{figure}
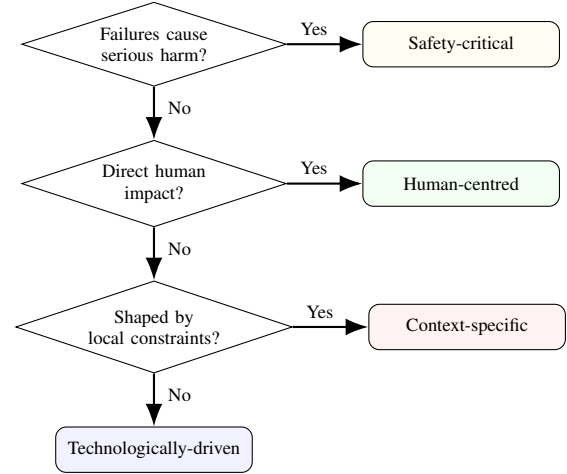

To assist developers in identifying the most relevant integration type, a lightweight decision flowchart (Figure~\ref{fig:integration_decision}) is provided. The flowchart guides designers through three high-level questions, ordered according to a risk-dependent logic:

\begin{enumerate}
    \item Will failures cause serious harm?
    \item Does the DT directly affect humans or communities?
    \item Is the system shaped primarily by local constraints?
\end{enumerate}

Based on the answers, the DT is categorised as one of the four categories as described in Section~\ref{sec:discussions}. This categorisation is intended as a guiding perspective and does not limit the DT's design, and every DT should consider all trust principles. Integration types help anticipate which trust concerns may require deeper analysis, more stringent controls, or earlier design decisions.

\begin{figure}
    \centering
    \includegraphics[width=7cm]{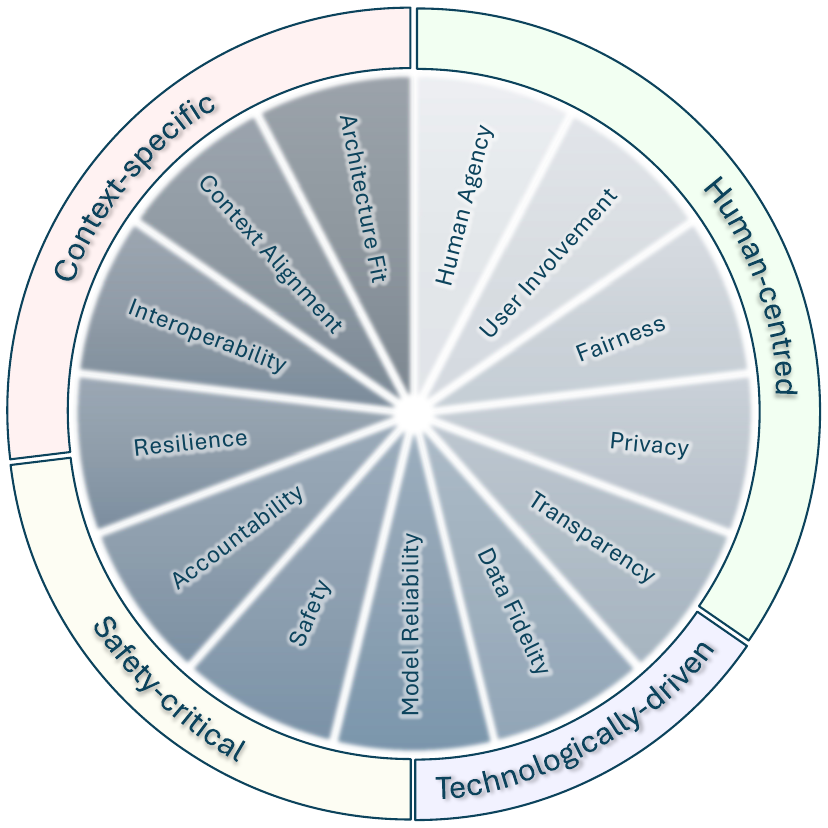}
    \caption{Mapping of trust-by-design principles to DT integration categories, indicating areas of higher emphasis across development stages.}
    \label{fig:trust_by_design_principle}
\end{figure}

To support prioritisation of trust considerations, 13 Trust-by-Design principles for DTs are defined, adapted from the Trust-by-Design framework proposed for collaborative intelligence systems \cite{Merchan2025trust} and reframed for explicit use during the design phase. Table~\ref{tab:trust_by_design_principle} summarises the principles and Figure~\ref{fig:trust_by_design_principle} highlights which integration categories they are particularly relevant to.

\begin{table*}[htbp]
\centering
\scriptsize
\caption{Trust-by-Design principles for DTs.} 
\label{tab:trust_by_design_principle} 
\begin{tabular}{p{2cm} p{10.5cm}}
\toprule
\textbf{Principle} & \textbf{Description} \\
\midrule
Human agency & Ensuring the DT supports, rather than replaces or obscures, human decision-making and oversight. \\ 
User involvement & Identifying relevant users and stakeholders early and integrating their requirements and constraints. \\ 
Fairness & Anticipating potential inequities or biases in data, models, or outcomes and designing mitigation. \\ 
Privacy & Incorporating privacy-by-design techniques before data pipelines are fixed. \\ 
Transparency & Clarifying modelling assumptions, data provenance, and system limitations from the outset. \\ 
Data fidelity & Planning for accurate, representative, and well-calibrated data sources. \\ 
Model reliability & Embedding validation, stress testing, and recalibration plans into the system specification. \\ 
Safety & Identifying potential harm pathways and integrating safety constraints or human authorisation. \\ 
Accountability & Defining responsibilities, governance roles, and escalation pathways before deployment. \\ 
Resilience & Designing for robustness, fault tolerance, and graceful degradation under adverse conditions. \\ 
Interoperability & Ensuring compatibility with data standards, regulatory requirements, and communication protocols. \\ 
Context alignment & Tailoring the DT to local workflows, environmental conditions, and domain-specific expectations. \\ 
Architecture fit & Selecting an appropriate level of decentralisation based on governance and operational constraints. \\ 
\bottomrule
\end{tabular} 
\end{table*}

Together, these principles form a practical design checklist, rather than a universal prescription. Future research should further refine how each principle affects different stakeholder groups, develop implementation patterns that make the principles actionable, and define measurable indicators to evaluate trust gains in deployed DT systems.

\subsection{Digital Twin as a Service (DTaaS) with Trust Metadata}

With DTs increasingly delivered as cloud-based or platform-based services, future research should further develop the concept of DTaaS. Existing DT modelling standards, such as the Digital Twin Definition Language (DTDL) from Microsoft Azure, the Next Generation Service Interfaces - Linked Data (NGSI-LD) developed by ETSI, and the Asset Administration Shell (AAS) from Germany's Industrie 4.0 initiative, provide structured schemas for describing entities, components, behaviours, and capabilities. However, these standards are primarily oriented toward functional characteristics and lack explicit mechanisms for representing trust-related attributes.

A major research opportunity lies in extending or harmonising DT schemas to incorporate trust metadata that can support assurance, transparency, and safe operation across DT ecosystems. Beyond attributes that are often discussed in the context of DT deployment, such as security guarantees, compliance requirements, transparency indicators, and provenance. There is scope to introduce richer metadata structures that can be encoded directly in DT schemas to enable trust-by-design. These include, but are not limited to:

\begin{itemize}
    \item Decision boundary metadata to specify when human oversight is required;
    \item Permission and role profiles to clarify who is authorised to control, observe, or modify components of a DT;
    \item Data quality indicators that communicate the expected reliability, freshness, or uncertainty of incoming data streams;
    \item Model verification audit trails that document validation procedures, stress tests, and update histories;
    \item Redundancy and fallback descriptions that declare failover models or backup data sources;
    \item Regulatory compliance tags to encode verifiable conformity with sector-specific standards; and
    \item Dependency graphs that capture both upstream and downstream DT dependencies, including reliance on external sensors, data pipelines, and other interconnected DTs.
\end{itemize}

An illustrative example of a DT model extended with trust metadata is shown in Figure~\ref{fig:dtdl_meta}, which presents a simplified HVAC system DTDL interface. The example demonstrates how decision boundaries, data quality indicators, redundancy information, regulatory tags, and dependency links could be incorporated directly into a DT schema.

\begin{figure}[ht]
\centering
\begin{minipage}{8.5cm}
\begin{lstlisting}[style=json]
{
  "@context": "dtmi:dtdl:context;2",
  "@id": "dtmi:example:HVAC;1",
  "@type": "Interface",
  "displayName": "HVAC System",
  "contents": [
    {
      "@type": "Telemetry",
      "name": "temperature",
      "schema": "double"
    },
    {
      "@type": "Property",
      "name": "targetTemperature",
      "schema": "double",
      "unit": "degreeCelsius"
    },
    {
      "@type": "Property",
      "name": "heatingEnabled",
      "schema": "boolean"
    }
  ],~
  "trust:metadata": {
    "decisionBoundary": "dtmi:example:HVAC_DecisionBoundary;1",
    "dataQualityIndicator": {
      "expectedCompleteness": 0.98,
      "expectedTimeliness": "500ms"
    },
    "redundancy": {
      "fallbackTwin": "dtmi:example:HVAC_Backup;1"
    },
    "regulatoryTags": ["ISO-27001", "GDPR-Art22"],
    "dependencyGraph": {
      "upstream": ["dtmi:example:Weather;1"],
      "downstream": ["dtmi:example:BMS;1"]
    }
  }~
}
\end{lstlisting}
\caption{Extended DTDL interface with trust metadata in line 24-38}
\label{fig:dtdl_meta}
\end{minipage}
\end{figure}

Embedding trust metadata into DT schemas would enable DTaaS platforms to expose their trust properties in a machine-interpretable and interoperable manner, improving transparency and accountability while providing clearer guidance to users when configuring, deploying, or procuring DT services. Future research and active industrial collaborations should focus on designing a comprehensive and interoperable set of trust metadata and exploring how DTaaS platforms could operationally respond to these attributes. For example, platforms could automatically select redundant or failover twins when a primary component fails, adjust system behaviour based on data quality or uncertainty metrics, log decision boundary events to enhance auditability and compliance, and propagate trust-related notifications or constraints to downstream DTs or users to support coordinated, trustworthy operation across interconnected DT ecosystems. Additional directions include investigating automated mechanisms to verify regulatory compliance, validate model assumptions, monitor interdependencies in real-time, and evaluate trade-offs between metadata complexity, interoperability, and system performance in large-scale deployments. Advancing these directions would help define what a trustworthy DT infrastructure should entail and operationalise trust-by-design principles in practice, providing users with clear, actionable information for managing DT services.

\subsection{Architectural Choices for Scalable and Trustworthy DTs}

Many works emphasise cloud-edge architectures or distributed DT designs, yet there is limited guidance on how to select architectural models that balance performance, accuracy, regulatory compliance, and system complexity. Implementing DTs can be costly, and both overly complex and overly simplistic models can affect user trust. Highly detailed or expansive DTs may require significant computational resources and maintenance costs, potentially reducing their perceived practical value, while simpler models may lack fidelity or reliability, undermining confidence in the system's outputs. These trade-offs underscore the need to carefully choose architectures that balance computational efficiency, model accuracy, operational robustness, and user expectations.

Centralised DTs offer a single, global view of the system, simplifying management and consistency, but they can be resource-intensive and may face privacy or regulatory challenges due to the aggregation of sensitive data. Distributed or edge-enabled DTs improve scalability, responsiveness, and resilience by deploying intelligence closer to physical assets, but they introduce additional risks, including cyber security vulnerabilities, potential inconsistencies, and challenges in maintaining synchronised data across multiple nodes.

An emerging and promising research direction is the concept of Federated Digital Twins (FDTs) \cite{Vergara2023federated}. An FDT consists of interconnected autonomous DTs, governed by rules for interoperability, coordination, and secure communications. Each DT maintains autonomy, supervises its own monitoring and virtual representation, and contributes to both individual and collective federation goals. FDTs enable cooperation, shared learning, and system-level simulations before transferring decisions to the physical entities. While the concept is still immature, it offers opportunities to explore how autonomous DTs can collaborate effectively, how trust metadata can be leveraged across federated environments, and how system-level performance, resilience, and regulatory compliance can be ensured. FDTs can adopt centralised, hierarchical, peer-to-peer, or regional architectural styles depending on system complexity, but the defining feature is the explicit governance of coordination and synchronisation among autonomous DTs.

Future research should systematically investigate architectural trade-offs across centralised, distributed, and federated paradigms. Key questions include how to partition modelling tasks, how much intelligence to push to the edge, and how architectural choices interact with trust, reliability, and socio-technical expectations. Evaluating FDTs in realistic scenarios will be particularly valuable to understand coordination mechanisms, interdependencies, synchronisation strategies, and communication overheads in large-scale, heterogeneous DT ecosystems. Advancing research in this area will provide practitioners with guidance for selecting architectures that are not only scalable and high-performing but also trustworthy, reliable, and aligned with practical user needs.

\section{Conclusion}
\label{sec:conclusion}

This study examines trustworthiness in DT systems, revealing that trustworthiness concerns are unevenly addressed across application domains. By distinguishing different deployment contexts, we clarify how trust priorities vary and identify opportunities for systematically embedding trust in DT design and operation. The findings provide a structured view of how trust is discussed across domains and offer actionable starting points for advancing the design of DT systems that are both technically robust and socially and operationally trustworthy. Three future research directions emerge from our analysis: embedding trust principles via Trust-by-Design frameworks, enhancing DT standards with trust metadata for interoperable DTaaS ecosystems, and evaluating architectural choices, including Federated DTs, to achieve scalable, reliable, and trustworthy deployments.

As DT technologies continue to evolve and integrate into critical infrastructures, addressing trustworthiness across technical, organisational, and human dimensions will be essential. Future research should focus on interdisciplinary approaches that operationalise these principles, ensuring that DT systems are not only powerful engineering tools but also transparent, accountable, and socially responsible technologies.

\bibliographystyle{ieeetr}
\bibliography{references}

\end{document}